Physics of Fluids

# Some general solutions for linear Bragg–Hawthorne equation

Cite as: Phys. Fluids **33**, 077113 (2021); https://doi.org/10.1063/5.0055228
Submitted: 27 April 2021 . Accepted: 19 June 2021 . Published Online: 16 July 2021
Ting Yi (蚁艇)

View Online   Export Citation   CrossMark

Physics of Fluids
SPECIAL TOPIC: Flow and Acoustics of Unmanned Vehicles
Submit Today!

AIP PublishingPhys. Fluids **33**, 077113 (2021); https://doi.org/10.1063/5.0055228    **33**, 077113
© 2021 Author(s).

# Some general solutions for linear Bragg–Hawthorne equation

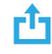

View Online    Export Citation    CrossMark

Ting Yi (蚁艇)[a]

**AFFILIATIONS**

Independent Researcher, Irvine, California 92617, USA

[a]Author to whom correspondence should be addressed: tingyi.physics@gmail.com

**ABSTRACT**

Linear cases of Bragg–Hawthorne equation for steady axisymmetric incompressible ideal flows are systematically discussed. The equation is converted to a more convenient form in a spherical coordinate system. A new vorticity decomposition is derived. General solutions for 16 linear cases of the equation are obtained. These solutions can be specified to gain new analytical vortex flows, as examples in the paper demonstrate. A lot of well-known solutions like potential flow past a sphere, Hill's vortex with and without swirl, are included and extended in these solutions. Special relations between some vortex flows are also revealed when exploring or comparing related solutions.

## I. INTRODUCTION

Bragg–Hawthorne equation[1] (or "B–H equation" in short) plays a central role in the study of axisymmetric steady flow of incompressible ideal fluids. The solutions of this equation (and its equivalent equation in magnetohydrodynamics, i.e., the Grad–Shafranov equation) can be applied to support a variety of research from hydrodynamics to magnetohydrodynamics and plasma physics, from vortex and turbulence study to rocket engine[2] and tokamak fusion reactor research,[3] from tornado study[4] to astrophysics,[5,6] etc.

Due to complexity of the equation, however, analytically solvable cases are rare. With special assumptions or restrictions for Bernoulli function and the azimuthal velocity, the equation reduces to a linear one. Most of the known analytic solutions so far are for such linear cases.

In this paper, an attempt is made to explore the linear cases in a systematic way and gain new explicit analytical solutions. First, the equation is rewritten in spherical coordinate system and the variable of polar angle is changed from $\theta$ to $\cos\theta$. This gives a new form of the equation that is more convenient to solve in most cases. Following that, a series of assumptions for Bernoulli function and the azimuthal velocity that make the equation linear are listed. These assumptions lead to 16 combinations, each associated with a special linear case of the equation. These 16 cases are then solved, mostly by separate variable method. The results are a series of general solutions, which, when specified, bring new analytical solutions of vortex flows. Many well-known solutions, including the potential flow around a sphere, Hill's vortex with and without swirl,[7,8] Bogoyavlenskij's solution of Beltrami flow,[9] the Fraenkel–Norbury family of vortex rings,[10,11] etc., can be obtained from these general solutions as well when constants are set to certain values.

Some linear cases of the equation have a close relation, so do their solutions and the related flows. For example, Hill's vortex without swirl is actually a combination of a uniform flow and an extra velocity. Hill's vortex with swirl can be considered as the Beltrami spherical vortex (the eigenvector field of curl operator in the spherical coordinate system, as discussed in Sec. VII) plus a special rotation. These are discussed in detail with their stream functions and velocities.

A new decomposition of vorticity is also derived. This decomposition applies to all flows (that follow B–H equation) and helps us to understand the impacts of Bernoulli function and the azimuthal velocity to vorticity and to the flows.

The rest of this paper is organized as follows: In Sec. II, we briefly recall the derivation of B–H equation, especially the form in spherical coordinate system. In Sec. III, series assumptions for Bernoulli function and azimuthal velocity are listed and briefly discussed. Section IV is about the vorticity decomposition. Sections V–X are devoted to solving different cases of the equation. Section XI offers a summary table and some further discussion.

## II. BRAGG–HAWTHORNE EQUATION IN SPHERICAL COORDINATES

For incompressible ideal fluids, the steady motions are described by (steady) Euler equations, which can be written as

$$\begin{cases} \nabla \cdot \mathbf{v} = 0, & (1) \\ \mathbf{v} \times (\nabla \times \mathbf{v}) = \nabla H, & (2) \end{cases}$$

where $\mathbf{v}$ is the velocity and $H$ is the Bernoulli function. LHS (left-hand side) of Eq. (2) is also known as the Lamb vector, denoted as

$$\mathbf{L} \equiv \mathbf{v} \times (\nabla \times \mathbf{v}) = \mathbf{v} \times \boldsymbol{\omega} \quad (\boldsymbol{\omega} \equiv \nabla \times \mathbf{v} \text{ is the vorticity}).$$

For Eq. (2) to stand, $\nabla \times \mathbf{L} = 0$ is required (and is sufficient, assuming that the flow is in a simply connected domain). A (non-irrotational) flow meeting this requirement is a generalized Beltrami flow[12] (or is further a Beltrami flow if $\mathbf{L} = 0$ in the whole domain[13]).

In the axisymmetric case, it is possible to define the Stokes stream function $\psi$ and an independent function $C$ so that

$$v_r = \frac{1}{r^2 \sin\theta} \frac{\partial \psi}{\partial \theta}, \quad v_\theta = -\frac{1}{r \sin\theta} \frac{\partial \psi}{\partial r}, \quad v_\varphi = \frac{C}{r \sin\theta} \quad (3)$$

in spherical coordinate system $(r, \theta, \varphi)$, or

$$v_\rho = -\frac{1}{\rho} \frac{\partial \psi}{\partial z}, \quad v_\phi = \frac{C}{\rho}, \quad v_z = \frac{1}{\rho} \frac{\partial \psi}{\partial \rho} \quad (4)$$

in cylindrical coordinate system $(\rho, \phi, z)$.

With the stream function, Eq. (1) is automatically satisfied. Taking the spherical coordinate case as an example, as all derivatives with respect to azimuthal angle $\varphi$ vanish, vorticity is

$$\boldsymbol{\omega} = \left(\frac{1}{r^2 \sin\theta} \frac{\partial C}{\partial \theta}\right) \mathbf{e}_r + \left(-\frac{1}{r \sin\theta} \frac{\partial C}{\partial r}\right) \mathbf{e}_\theta$$
$$+ \left(-\frac{1}{r \sin\theta} \frac{\partial^2 \psi}{\partial r^2} + \frac{\cos\theta}{r^3 \sin^2\theta} \frac{\partial \psi}{\partial \theta} - \frac{1}{r^3 \sin\theta} \frac{\partial^2 \psi}{\partial \theta^2}\right) \mathbf{e}_\varphi \quad (5)$$

and the three components of Lamb vector are

$$\begin{cases} L_r = \left(-\frac{1}{r \sin\theta} \frac{\partial \psi}{\partial r}\right) \left(-\frac{1}{r \sin\theta} \frac{\partial^2 \psi}{\partial r^2} + \frac{\cos\theta}{r^3 \sin^2\theta} \frac{\partial \psi}{\partial \theta} - \frac{1}{r^3 \sin\theta} \frac{\partial^2 \psi}{\partial \theta^2}\right) \\ \qquad - \left(\frac{C}{r \sin\theta}\right) \left(-\frac{1}{r \sin\theta} \frac{\partial C}{\partial r}\right), & (6) \\ L_\theta = \frac{C}{r \sin\theta} \left(\frac{1}{r^2 \sin\theta} \frac{\partial C}{\partial \theta}\right) - \frac{1}{r^2 \sin\theta} \frac{\partial \psi}{\partial \theta} \\ \qquad \times \left(-\frac{1}{r \sin\theta} \frac{\partial^2 \psi}{\partial r^2} + \frac{\cos\theta}{r^3 \sin^2\theta} \frac{\partial \psi}{\partial \theta} - \frac{1}{r^3 \sin\theta} \frac{\partial^2 \psi}{\partial \theta^2}\right), & (7) \\ L_\varphi = \frac{1}{r^3 \sin^2\theta} \left(\frac{\partial \psi}{\partial r} \frac{\partial C}{\partial \theta} - \frac{\partial \psi}{\partial \theta} \frac{\partial C}{\partial r}\right). & (8) \end{cases}$$

As the flow is steady, each of $H$ and $C$ is a function of $\psi$ only.[14] Thus, we have the following "chain rules" for $H(\psi)$ and $C(\psi)$:

$$\frac{\partial H}{\partial r} = \frac{dH}{d\psi} \frac{\partial \psi}{\partial r}, \quad \frac{\partial H}{\partial \theta} = \frac{dH}{d\psi} \frac{\partial \psi}{\partial \theta}, \quad \frac{\partial C}{\partial r} = \frac{dC}{d\psi} \frac{\partial \psi}{\partial r}, \quad \frac{\partial C}{\partial \theta} = \frac{dC}{d\psi} \frac{\partial \psi}{\partial \theta}. \quad (9)$$

In this case, the Lamb vector components in Eqs. (6)–(8) simplify to

$$\begin{cases} L_r = \frac{1}{r^2 \sin^2\theta} \left(\frac{\partial^2 \psi}{\partial r^2} - \frac{\cos\theta}{r^2 \sin\theta} \frac{\partial \psi}{\partial \theta} + \frac{1}{r^2} \frac{\partial^2 \psi}{\partial \theta^2} + C \frac{dC}{d\psi}\right) \frac{\partial \psi}{\partial r}, & (10) \\ L_\theta = \frac{1}{r^3 \sin^2\theta} \left(\frac{\partial^2 \psi}{\partial r^2} - \frac{\cos\theta}{r^2 \sin\theta} \frac{\partial \psi}{\partial \theta} + \frac{1}{r^2} \frac{\partial^2 \psi}{\partial \theta^2} + C \frac{dC}{d\psi}\right) \frac{\partial \psi}{\partial \theta}, & (11) \\ L_\varphi = 0. & (12) \end{cases}$$

Equation (2), if rewritten with Eq. (9) and Eqs. (10)–(12), becomes an identity in $\mathbf{e}_\varphi$ direction (both sides vanish) and two identical scalar equations in $\mathbf{e}_r$ and $\mathbf{e}_\theta$ direction, and both read

$$\frac{\partial^2 \psi}{\partial r^2} - \frac{\cos\theta}{r^2 \sin\theta} \frac{\partial \psi}{\partial \theta} + \frac{1}{r^2} \frac{\partial^2 \psi}{\partial \theta^2} = r^2 \sin^2\theta \frac{dH}{d\psi} - C \frac{dC}{d\psi}. \quad (13)$$

This is the Bragg–Hawthorne equation[1] (in spherical coordinate system). It is also referred to as Hicks equation[15] or Squire–Long equation.[16,17] The Grad–Shafranov equation in ideal magnetohydrodynamics also has the same form.[18,19]

Changing the variable from $\theta$ to $\cos\theta$, we have $\frac{\partial \psi}{\partial \theta} = -\sin\theta \frac{\partial \psi}{\partial (\cos\theta)}$ and $\frac{\partial^2 \psi}{\partial \theta^2} = \sin^2\theta \frac{\partial^2 \psi}{\partial (\cos\theta)^2} - \cos\theta \frac{\partial \psi}{\partial (\cos\theta)}$. Equation (13) becomes

$$\frac{\partial^2 \psi}{\partial r^2} + \frac{\sin^2\theta}{r^2} \frac{\partial^2 \psi}{\partial (\cos\theta)^2} = r^2 \sin^2\theta \frac{dH}{d\psi} - C \frac{dC}{d\psi}. \quad (14)$$

Compared to Eq. (13), LHS of Eq. (14) only has two terms. Considering $\sin^2\theta = 1 - \cos^2\theta$, Eq. (14) has a simple form with respect to the variable $\cos\theta$. This is convenient. A lot of solutions in this paper are based on this form of the equation.

It is worthwhile to point out that a similar form of Grad–Shafranov equation (using $\cos\theta$ as the variable) is presented by Low and Lou (1990) in a solar magnetic field study paper.[5]

The above derivation of B–H equation also applies to the case in cylindrical coordinate system $(\rho, \phi, z)$. In that, Eq. (13) or (14) takes the well-known form as

$$\frac{\partial^2 \psi}{\partial \rho^2} - \frac{1}{\rho} \frac{\partial \psi}{\partial \rho} + \frac{\partial^2 \psi}{\partial z^2} = \rho^2 \frac{dH}{d\psi} - C \frac{dC}{d\psi}. \quad (15)$$

This equation, either in the form of Eqs. (13), (14), or (15), is the governing equation for ideal incompressible axisymmetric steady flows. For a specific case of the functions $H(\psi)$ and $C(\psi)$, if we can find the stream function $\psi$ satisfying the equation, velocity can be calculated and Euler equation (2) is solved.

In general, however, this could be difficult. Mathematically, $H(\psi)$ and $C(\psi)$ can be arbitrary (smooth) functions of $\psi$. They can make the equation complicated. Nevertheless, when $H(\psi)$ and $C(\psi)$ are in certain forms, the equation can be linear and solvable.

### III. LINEAR CASES OF BRAGG–HAWTHORNE EQUATION

LHS of the equation, either in Eqs. (13), (14), or (15), is linear with respect to the stream function $\psi$. If the two terms on the RHS are also linear, the equation is linear.

Assuming $H_0, \psi_0, \lambda, \lambda_1, \lambda_2, a$ and $a_0, a_1, a_2, a_3$ are constants, for the first term on RHS to be linear, we have

($H_1$): $H = H_0$ (then $r^2 \sin^2\theta \frac{dH}{d\psi} = 0$, $\rho^2 \frac{dH}{d\psi} = 0$),

($H_2$): $H = H_0 + \lambda \psi$ (then $r^2 \sin^2\theta \frac{dH}{d\psi} = \lambda r^2 \sin^2\theta$, $\rho^2 \frac{dH}{d\psi} = \lambda \rho^2$),

($H_3$): $H = H_0 + \lambda \psi^2$ (then $r^2 \sin^2\theta \frac{dH}{d\psi} = 2\lambda r^2 \sin^2\theta \psi$, $\rho^2 \frac{dH}{d\psi} = 2\lambda \rho^2 \psi$),

($H_4$): $H = H_0 + \lambda_1 \psi + \lambda_2 \psi^2$ (then $r^2 \sin^2\theta \frac{dH}{d\psi} = 2\lambda_2 r^2 \sin^2\theta \psi + \lambda_1 r^2 \sin^2\theta$, $\rho^2 \frac{dH}{d\psi} = 2\lambda_2 \rho^2 \psi + \lambda_1 \rho^2$).

For the second term to be linear, we can have

($C_1$): $C = a\psi_0$ (then $C \frac{dC}{d\psi} = 0$),

($C_2$): $C = a_1 \sqrt{\psi_0 + a_2 \psi}$ (then $C \frac{dC}{d\psi} = \frac{1}{2} a_1^2 a_2 = \frac{1}{2} a$, re-denoting $a \equiv a_1^2 a_2$),

($C_3$): $C = \psi_0 + a\psi$ (then $C \frac{dC}{d\psi} = a^2 \psi + a\psi_0$),

($C_4$): $C = a_1 \sqrt{\psi_0^2 + a_2 \psi + a_3 \psi^2}$ (then $C \frac{dC}{d\psi} = a_1^2 a_3 \psi + \frac{1}{2} a_1^2 a_2 = a\psi + a_0$, re-denoting $a \equiv a_1^2 a_3$, $a_0 \equiv \frac{1}{2} a_1^2 a_2$).

Mathematically, some constants in the list can be absolved or combined. However, keeping them in the above format will make it convenient for discussion.

Constant $\psi_0$ in $C_1$, $C_2$, and $C_4$ does not impact the equation and thus has no impact on the solutions. However, when $C$ is interpreted linking to azimuthal velocity, $\psi_0$ can bring big differences for azimuthal velocity and for the flow. For $C_3$, most of the feasible cases require $\psi_0 = 0$. However, to maximize generality, $\psi_0$ is kept here as a constant, and the different impacts of $\psi_0 = 0$ and $\psi_0 \neq 0$ will be discussed specifically.

The cases in the list are not always independent. For example, $H_1$ can be considered as a special case of $H_2$ (when $\lambda = 0$); $H_4$ is a "combination" of $H_2$ and $H_3$. However, the equations and solutions for these cases are quite different and are worthy to be discussed separately. For this consideration, they are listed as separated cases.

$C_3$ and $C_4$ actually give the same form of equation (with different denotations for the constants). So, they would share the same stream function as mathematical solution of the equation. However, when the same solution is respectively interpreted for $C_3$ and $C_4$, the flows can be significantly different (on velocity, vorticity, physical feasibility, etc.). For completeness consideration, they are treated as two separated cases in the list.

In a more general view, for linear cases, when the two terms on RHS are expanded, they can become up to four terms, which can be written on RHS of the following "model equation" as

$$\frac{\partial^2 \psi}{\partial r^2} + \frac{\sin^2\theta}{r^2} \frac{\partial^2 \psi}{\partial (\cos\theta)^2} = \Lambda_1 r^2 \sin^2\theta \psi + \Lambda_0 r^2 \sin^2\theta - A_1 \psi - A_0 \tag{16}$$

or in cylindrical coordinates

$$\frac{\partial^2 \psi}{\partial \rho^2} - \frac{1}{\rho} \frac{\partial \psi}{\partial \rho} + \frac{\partial^2 \psi}{\partial z^2} = \Lambda_1 \rho^2 \psi + \Lambda_0 \rho^2 - A_1 \psi - A_0. \tag{17}$$

$\Lambda_1$, $\Lambda_0$, $A_1$, and $A_0$ here represent certain form of the constants in the list. Different cases of function $H$ and $C$ are associated with different combinations of presentation or absence of these four terms.

Obviously, with $H_1$–$H_4$ and $C_1$–$C_4$, there are 16 combinations, each making Eqs. (13), (14), or (15) a linear equation of $\psi$ (i.e., each is related to a combination of the $\Lambda_1$, $\Lambda_0$, $A_1$, and $A_0$ terms in the model equation). For convenience, we will denote these 16 combinations as $H_m C_n$ (m, n = 1, 2, 3, 4) going forward in this paper.

## IV. VORTICITY DECOMPOSITION AND FLOW PROPERTIES

Before exploring solutions of particular cases, it is worthy to further investigate impacts of the two functions, $H$ and $C$, to vorticity and to the flow.

Recalling Eqs. (3) and (4), we have $v_\varphi = \frac{C}{r\sin\theta}$ (or $v_\phi = \frac{C}{\rho}$ in cylindrical coordinates). Function $C$ is indeed the linear azimuthal velocity multiplied by distance to the z-axis. $2\pi C$ is often considered as circulation along a circle around z-axis. From another angle of view, $C$ can also be considered as the "azimuthal velocity moment" with respect to z-axis. As mass density is a constant in incompressible fluid, $C$ is also representing angular momentum density of the fluid (with respect to z-axis).

In any of these considerations, an assumption (or a restriction) on $C$ is basically an assumption (restriction) on azimuthal velocity related to the stream function $\psi$. In other words, $C$ indicates how the fluid is moving around the symmetric axis. It is intuitive to expect that $C$ has close relation with vorticity of the flow.

Bernoulli function $H$, on the other hand, is the inverse gradient of Lamb vector, which is just the cross-product of velocity and vorticity. It is also natural to expect that $H$ is closely related to vorticity.

With Eqs. (3), (9), and (13), vorticity in Eq. (5) can be rewritten as

$$\boldsymbol{\omega} = \frac{dC}{d\psi} \left( \frac{1}{r^2 \sin\theta} \frac{\partial \psi}{\partial \theta} \boldsymbol{e}_r - \frac{1}{r \sin\theta} \frac{\partial \psi}{\partial r} \boldsymbol{e}_\theta + \frac{C}{r \sin\theta} \boldsymbol{e}_\varphi \right) - r \sin\theta \frac{dH}{d\psi} \boldsymbol{e}_\varphi,$$

which is also

$$\boldsymbol{\omega} = \frac{dC}{d\psi} \boldsymbol{v} - r \sin\theta \frac{dH}{d\psi} \boldsymbol{e}_\varphi. \tag{18}$$

Same relation exists as well in cylindrical coordinates, where we have

$$\boldsymbol{\omega} = \frac{dC}{d\psi} \boldsymbol{v} - \rho \frac{dH}{d\psi} \boldsymbol{e}_\phi. \tag{19}$$

Either Eq. (18) or Eq. (19) indicates that vorticity $\boldsymbol{\omega}$ can be split into two portions. One is in the velocity direction, proportional to velocity with a scale factor $\frac{dC}{d\psi}$. The second portion is on the azimuthal direction, with magnitude proportional to $\frac{dH}{d\psi}$ (and to the distance to z-axis).

In such a sense, it can be considered that functions $H$ and $C$ construct vorticity, or more specifically, derivatives of $H$ and $C$ compose vorticity in the flow. $\frac{dC}{d\psi}$ decides vorticity in the direction of velocity; $\frac{dH}{d\psi}$ decides vorticity in azimuthal direction. Although these two directions in general are not orthogonal to each other, these two components make up the vorticity at each point in the flow.

Equation (18) or Eq. (19) thus can be considered as a decomposition of vorticity. It applies to all flows that B–H equation stands (i.e., all axisymmetric steady flows of incompressible ideal fluid).

Actually, the relations of $H$ and $C$ to the vorticity components are obvious and are often implied in derivation of equations (see, e.g., Batchelor,[14] Sec. 7.5). However, explicitly putting them in the form of Eq. (18) or Eq. (19) straightforwardly shows the impacts of $H$ and $C$ to vorticity and to the flow.

For convenience, we will further denote $\boldsymbol{\omega_B} \equiv \frac{dC}{d\psi}\boldsymbol{v}$, $\boldsymbol{\omega_A} \equiv -r\sin\theta\frac{dH}{d\psi}\boldsymbol{e_\varphi}$ (or $\boldsymbol{\omega_A} \equiv -\rho\frac{dH}{d\psi}\boldsymbol{e_\phi}$ in cylindrical coordinates) and call $\boldsymbol{\omega_B}$ and $\boldsymbol{\omega_A}$ Beltrami vorticity and azimuthal vorticity, respectively. Total vorticity then is sum of these two vectors, i.e., $\boldsymbol{\omega} = \boldsymbol{\omega_B} + \boldsymbol{\omega_A}$.

As Beltrami vorticity $\boldsymbol{\omega_B}$ is parallel to velocity, it does not impact Lamb vector. We can just count on the azimuthal vorticity $\boldsymbol{\omega_A}$ when calculating Lamb vector. In other words, regardless of $\boldsymbol{\omega_B}$, we always have $\boldsymbol{L} = \boldsymbol{v} \times \boldsymbol{\omega_A} = -r\sin\theta\frac{dH}{d\psi}\boldsymbol{v} \times \boldsymbol{e_\varphi}$ (or $\boldsymbol{L} = -\rho\frac{dH}{d\psi}\boldsymbol{v} \times \boldsymbol{e_\phi}$ in cylindrical coordinates).

This is consistent with the fact that (in ideal incompressible axisymmetric steady flows) Lamb vector is always perpendicular to the azimuthal direction (except for points in the symmetric axis). Following this is also the corollary that Lamb vector is always coplanar with the symmetric axis in such flows.

Applying this decomposition to flows in the list in Sec. III, for the $H_1C_n$ cases, as $\frac{dH}{d\psi} = 0$, azimuthal vorticity $\boldsymbol{\omega_A}$ vanishes. Total vorticity contains solely the Beltrami part $\boldsymbol{\omega_B}$. Lamb vector thus vanishes as well. For $H_1C_1$, $\boldsymbol{\omega_B}$ also vanishes, so does the total vorticity. It is hence an irrotational/potential axisymmetric flow. For $H_1C_2$, $\boldsymbol{\omega} = \boldsymbol{\omega_B} = \frac{a_1 a_2}{2\sqrt{\psi_0 + a_2\psi}}\boldsymbol{v}$. Vorticity is parallel to velocity with a non-constant coefficient. This is a non-linear Beltrami flow (i.e., the velocity field is a non-linear Beltrami field). For $H_1C_3$, we have $\boldsymbol{\omega} = \boldsymbol{\omega_B} = \alpha\boldsymbol{v}$. Vorticity is parallel to velocity with a constant coefficient. This is a linear Beltrami case. $H_1C_4$, on the other hand, is a non-linear Beltrami flow again, as $\boldsymbol{\omega} = \boldsymbol{\omega_B} = \frac{a_1(a_2 + 2a_3\psi)}{2\sqrt{\psi_0^2 + a_2\psi + a_3\psi^2}}\boldsymbol{v}$.

In the $H_2C_n$ family, azimuthal vorticity $\boldsymbol{\omega_A} = -\lambda r\sin\theta\boldsymbol{e_\varphi}$ (or $\boldsymbol{\omega_A} = -\lambda\rho\boldsymbol{e_\phi}$ in cylindrical coordinates). It is proportional to the distance to z-axis. For $H_2C_1$, $\boldsymbol{\omega_B}$ vanishes. Total vorticity is made up solely by $\boldsymbol{\omega_A}$ and thus is proportional to the distance to z-axis as well. This is actually the assumption of the Fraenkel–Norbury family of vortex rings[10,11] (including the Hill's vortex without swirl). As can be expected, the general solution of $H_2C_1$ will include Hill's vortex as a special case. $H_2C_3$ is adding the Beltrami vorticity part $\boldsymbol{\omega_B} = \alpha\boldsymbol{v}$ to $H_2C_1$. This is actually the case of Hill's vortex with swirl, which will be discussed in Sec. IX. For $H_2C_2$ and $H_2C_4$, both Beltrami vorticity $\boldsymbol{\omega_B}$ and azimuthal vorticity $\boldsymbol{\omega_A}$ present in the total vorticity, and the flows have more complicated dependency to $H$ and $C$.

In the $H_3C_n$ and $H_4C_n$ cases, as $\frac{dH}{d\psi}$ explicitly contains $\psi$, azimuthal vorticity is now directly tied with $\psi$ (in opposite to $H_1C_n$ or $H_2C_n$ where $\psi$ is not explicitly involved in the expression of azimuthal vorticity). The equation is still linear, but this azimuthal vorticity brings an extra term with the unknown function $\psi$ to the equation and makes the solutions complicated. These cases will be further discussed in Sec. X.

As an aside, the second term on RHS of the B–H equation, $C\frac{dC}{d\psi}$, has a notable property of symmetric. This can also be analyzed by applying the vorticity decomposition.

Mathematically, function $C$ can change to the opposite sign without impacting the equation. If $\psi$ is a solution of the B–H equation related to $C(\psi)$, $\psi$ is also a solution of the equation related to $-C(\psi)$. Physically, this implies that flows governed by B–H equation are "two-way flows." The azimuthal velocity $v_\varphi = \frac{C}{r\sin\theta}$ or $v_\phi = \frac{C}{\rho}$ can change to the opposite direction and the flow still satisfies the same equation. Equivalently speaking, for each flow as a solution of B–H equation, there exists its chiral symmetric flow (with opposite azimuthal velocity) to form a pair.

From vorticity decomposition point of view, when $C(\psi)$ changes sign (and $\psi$ remains the same), velocity

$$\boldsymbol{v} = \frac{1}{r^2\sin\theta}\frac{\partial\psi}{\partial\theta}\boldsymbol{e_r} - \frac{1}{r\sin\theta}\frac{\partial\psi}{\partial r}\boldsymbol{e_\theta} + \frac{C}{r\sin\theta}\boldsymbol{e_\varphi}$$

changes chirality [i.e., from $(v_r, v_\theta, v_\varphi)$ to $(v_r, v_\theta, -v_\varphi)$], but Beltrami vorticity

$$\boldsymbol{\omega_B} = \frac{dC}{d\psi}\left(\frac{1}{r^2\sin\theta}\frac{\partial\psi}{\partial\theta}\boldsymbol{e_r} - \frac{1}{r\sin\theta}\frac{\partial\psi}{\partial r}\boldsymbol{e_\theta} + \frac{C}{r\sin\theta}\boldsymbol{e_\varphi}\right)$$

undergoes a half-round rotation along $\boldsymbol{e_\varphi}$ direction without changing chirality [i.e., from $(\omega_{Br}, \omega_{B\theta}, \omega_{B\varphi})$ to $(-\omega_{Br}, -\omega_{B\theta}, \omega_{B\varphi})$]. At the same time, azimuthal vorticity $\boldsymbol{\omega_A} = -r\sin\theta\frac{dH}{d\psi}\boldsymbol{e_\varphi}$ remains unchanged. Total vorticity $\boldsymbol{\omega} = \boldsymbol{\omega_B} + \boldsymbol{\omega_A}$ thus changes to a new vector that, in general, is neither rotational symmetric nor chirally symmetric to the original vorticity before $C$ changes sign. Despite the change on vorticity, however, Lamb vector remains the same, and so does the equation (and the solution).

## V. AXISYMMETRIC POTENTIAL FLOW ($H_1C_1$)

Combination $H_1C_1$ is the simplest one in the list. As discussed in Sec. IV, all flows in this case are axisymmetric potential flow.

As both terms on RHS of B–H equation vanish, Eq. (14) becomes

$$\frac{\partial^2\psi}{\partial r^2} + \frac{\sin^2\theta}{r^2}\frac{\partial^2\psi}{\partial(\cos\theta)^2} = 0. \quad (20)$$

Let $\psi = R(r)\Theta(\cos\theta)$ and apply variable separation method. With the separation constant denoted as $n(n + 1)$, the two separated questions are

$$r^2\ddot{R} - n(n+1)R = 0, \quad (21)$$

$$\sin^2\theta\ddot{\Theta} + n(n+1)\Theta = 0. \quad (22)$$

Solution of Eq. (21) is $R = k_1 r^{n+1} + k_2 r^{-n}$.

$k_1, k_2, n$ here, as well as $k_3, k_4, k_5, k_6, k_0, k,$ and $K_n$ in expressions later in this paper are independent constants. Mathematically, they can be arbitrary real numbers, or theoretically even be imaginary numbers in some cases. However, to have the stream function and velocity physically meaningful, restrictions may apply to them.

In this paper, we will consider a flow (from a solution of the B–H equation) with no singular velocity (i.e., no infinite nor discontinuous velocity) in a domain as "physically feasible." When the domain is not the whole space, the boundary conditions can be provided by a solid boundary or by other flows outside the domain. In the latter case, the flows inside and outside the domain should have zero normal velocity and continuous tangential velocity at the boundary. Equivalently, this requires inside and outside flows to have constant (normally zero) stream function and continuous stream function gradient on the boundary.

For Eq. (22), $\Theta$ is a function of $\cos\theta$ and derivative $\ddot{\Theta}$ is with respect to $\cos\theta$. If we replace the unknown function $\Theta(\cos\theta)$ with $\sin\theta T(\cos\theta)$ and apply the identities $\frac{d(\sin\theta)}{d(\cos\theta)} = -\frac{\cos\theta}{\sin\theta}, \frac{d^2(\sin\theta)}{d(\cos\theta)^2} = -\frac{1}{\sin^3\theta}$, Eq. (22) becomes $\sin^2\theta\ddot{T} - 2\cos\theta\dot{T} + \left[n(n+1) - \frac{1}{\sin^2\theta}\right]T = 0$. This is an associated Legendre equation of $T(\cos\theta)$ and thus is solved by linear combination of the associated Legendre function of the first and second kind, $P_n^1(\cos\theta)$ and $Q_n^1(\cos\theta)$. Solution of Eq. (22) hence can be written as

$$\Theta(\cos\theta) = \sin\theta \left[ k_3 P_n^1(\cos\theta) + k_4 Q_n^1(\cos\theta) \right]. \quad (23)$$

Combining it with solution of Eq. (21) gives the general solution of Eq. (20) as

$$\psi = \left( k_1 r^{n+1} + k_2 r^{-n} \right) \sin\theta \left[ k_3 P_n^1(\cos\theta) + k_4 Q_n^1(\cos\theta) \right]. \quad (24)$$

With special settings for the constants, this solution can give specific solutions in simple forms. For example, when $n = 1$, $k_2 = 0$, $k_3 = -1$, $k_4 = 0$, Eq. (24) becomes $\psi = k_1 r^2 \sin^2\theta$. This is a uniform flow with velocity $U = 2k_1$ in z-direction.

If $k_1$ and $k_2$ are both non-zero, the radial portion in Eq. (24), $R(r) = k_1 r^{n+1} + k_2 r^{-n}$ has a zero at $r = r_0 = \left( -\frac{k_2}{k_1} \right)^{\frac{1}{2n+1}}$. In the case that such $r_0$ is a positive real number, stream function $\psi$ is zero at the surface of the sphere $r = r_0$. The flow outside the sphere is a potential flow around that sphere (while the flow inside the sphere is singular at $r = 0$ and thus is not physically feasible).

In this case, if we further have $n = 1$, $k_4 = 0$, we have $\psi = \frac{U}{2} r^2 \left(1 - \frac{a^3}{r^3}\right) \sin^2\theta$ (where $U = -2k_1 k_3$, $a = r_0 = \left(-\frac{k_2}{k_1}\right)^{1/3}$). This is the well-known potential flow past the sphere $r = a$.

When $n > 1$, the associated Legendre functions in Eq. (24) can give more complicated potential flows around the sphere. Some examples can be found in Sec. VII [i.e., Eq. (37b) and Fig. 5].

As a property of the $C_1$ flows (i.e., flows in any $H_mC_1$ case), as long as function $C$ is a constant, its value does not impact the equation and thus will not impact the solution. Azimuthal velocity $v_\varphi = \frac{C}{r\sin\theta}$ hence is independent of the solution. In principle, velocity $(v_r, v_\theta)$ in the meridian plane derived from solution of an $H_mC_1$ equation can be with any azimuthal velocity $v_\varphi = \frac{C}{r\sin\theta}$ as long as $C$ is a constant. When $C \neq 0$, this $v_\varphi$ introduces a circular movement with uniform angular momentum density with respect to z-axis. This is an "irrotational rotation" (as it does not impact the vorticity). In this sense, constant $C$ brings a rotation to $H_mC_1$ flows without impacting vorticity nor impacting velocity in the meridian plane.

Such a rotation has singular velocity at z-axis, though. It is physically feasible only for special cases (e.g., in a domain that is not overlapping z-axis. An example on $H_2C_1$ can be found in Sec. VIII, which is the swirled case of Fraenkel–Norbury solutions.).

Case $H_1C_1$ can also be solved in cylindrical coordinates. Equation (15) for $H_1C_1$ is

$$\frac{\partial^2 \psi}{\partial \rho^2} - \frac{1}{\rho} \frac{\partial \psi}{\partial \rho} + \frac{\partial^2 \psi}{\partial z^2} = 0. \quad (25)$$

Let $\psi(\rho, z) = P(\rho) Z(z)$, Eq. (25) can be separated to two equations as $\rho \ddot{P} - \dot{P} + k\rho P = 0$ and $\ddot{Z} - kZ = 0$ ($k$ is the separation constant). The latter one is solved by $Z(z) = k_3 e^{\sqrt{k}z} + k_4 e^{-\sqrt{k}z}$. The former one can be converted to a Bessel equation of $F(x)$ (of order 1) by replacing $P(\rho)$ by $xF(x)$ (with $x = \sqrt{k}\rho$) and thus is solved by linear combination of the Bessel function of the first and second kind, $J_1$ and $Y_1$. Mathematically, the general solution for Eq. (25) is

$$\psi = \left[ k_1 \rho J_1(\sqrt{k}\rho) + k_2 \rho Y_1(\sqrt{k}\rho) \right] \left( k_3 e^{\sqrt{k}z} + k_4 e^{-\sqrt{k}z} \right). \quad (26)$$

This solution is actually obtained and published previously (1963) by Berker.[20]

When $k > 0$, $Z(z) = (k_3 e^{\sqrt{k}z} + k_4 e^{-\sqrt{k}z})$ is unbounded when $z \to +\infty$ and/or $z \to -\infty$, and thus a feasible flow can only exist in a domain excluding upper or lower "end" of the z-axis. When $k < 0$, with proper $k_3$ and $k_4$, $Z(z)$ is periodic (and bounded) along the whole z-axis. (Alternatively, in this case equation $\ddot{Z} - kZ = 0$ can be solved by $Z(z) = k_3 \sin(\sqrt{-k}z) + k_4 \cos(\sqrt{-k}z)$, which is equivalent to the above solution in exponential function form, and is periodic along z-axis.). However, in this case, $\sqrt{k}$ is not a real number, and the two Bessel functions in Eq. (26) become modified Bessel functions (of the first and second kind, respectively), which are unbounded on z-axis (i.e., $\rho = 0$) or when $\rho$ approaches infinity.

Overall, physical feasibility of Eq. (26) is limited and needs to be carefully considered according to domain of the flow and boundary conditions, etc.

In passing, $\psi = k_1 \rho^2 + k_2 z + k_3$ is also a solution of Eq. (25). This is not from variable separation method and thus is not included in Eq. (26). When $k_2 = 0$ and $k_3 = 0$, it converts to $\psi = k_1 \rho^2$, which is equivalent to the special form of Eq. (24), i.e., the uniform flow $\psi = k_1 r^2 \sin^2\theta$.

## VI. AXISYMMETRIC NON-LINEAR BELTRAMI FLOW ($H_1C_2$)

The solution of $H_1C_1$ can be "extended" to solve $H_1C_2$. Taking the spherical coordinates case first, Eq. (14) for $H_1C_2$ is

$$\frac{\partial^2 \psi}{\partial r^2} + \frac{\sin^2\theta}{r^2} \frac{\partial^2 \psi}{\partial (\cos\theta)^2} = -\frac{a}{2}. \quad (27)$$

This can be considered as the homogeneous equation Eq. (20) plus a constant inhomogeneous term on the RHS. It thus can be solved by adding a particular solution, $\psi^* = -\frac{a}{4} r^2$ to general solution of Eq. (20). This gives

$$\psi = \left( k_1 r^{n+1} + k_2 r^{-n} \right) \sin\theta \left[ k_3 P_n^1(\cos\theta) + k_4 Q_n^1(\cos\theta) \right] - \frac{a}{4} r^2. \quad (28)$$

Compared to Eq. (24), the particular solution $\psi^*$ brings new features for the flow of Eq. (28). It actually turns the potential flow in Eq. (24) into a non-linear Beltrami flow.

If we calculate the velocity from Eq. (28) and compare it with velocity from Eq. (24), $\psi^*$ does not impact $v_r$, but it brings an addition $v_\theta^* = \frac{a}{2\sin\theta}$ to the polar velocity $v_\theta$. In the meridian plane, this $v_\theta^*$ represents a "rotation" around the origin point, with different magnitudes at different polar angles (and is singular on z-axis). Figure 1 shows this $v_\theta^*$ in x–z plane (when $a = 1$).

The impact of $\psi^*$ to azimuthal velocity is more complicated. As $v_\varphi = \frac{a_1 \sqrt{\psi_0 + a_2 \psi}}{r\sin\theta}$ is a non-linear function of $\psi$, the impact of $\psi^*$ to $v_\varphi$ is not simply a linear addition.

Theoretically, Eq. (28) can be specified to various flows depending on the constants. However, most (if not all) of them may inherit the singularity from $v_\theta^*$ and thus need to be within a domain excluding z-axis. Besides, for $v_\varphi$ to be real, it is required that $\psi_0 + a_2\psi \geq 0$. This may add more restrictions to the flow.

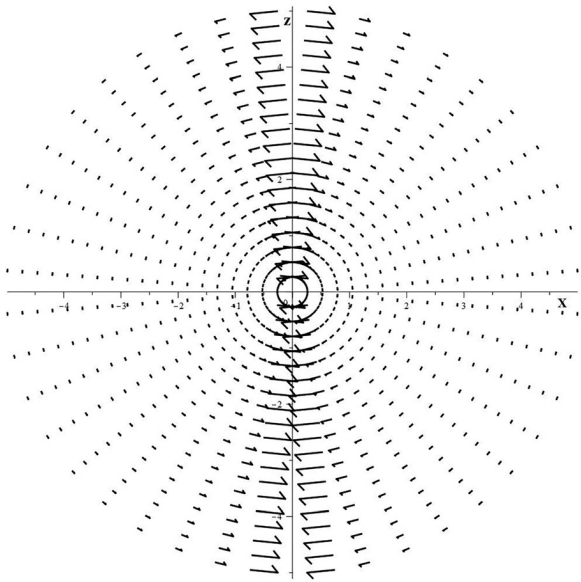

FIG. 1. $v_\theta^*$ in x–z plane brought by $\psi^*$ for $H_1C_2$ flow.

As an example of the feasible cases, when $k_1 = \frac{a}{2}$ and $n = 1$, $k_2 = 0$, $k_3 = -1$, $k_4 = 0$, stream function (28) becomes $\psi = -\frac{a}{4}r^2\cos(2\theta)$. This is a combination of the uniform flow of Eq. (24) with $U = a$ in z-direction (i.e., $\psi = \frac{a}{2}r^2\sin^2\theta$) and the rotation of $v_\theta^*$ shown in Fig. 1 (that is brought by the particular solution $\psi^* = -\frac{a}{4}r^2$).

Velocity in this case is $v_r = a\cos\theta$, $v_\theta = \frac{a}{2}\frac{\cos(2\theta)}{\sin\theta}$, $v_\varphi = \frac{a_1\sqrt{\psi_0 - \frac{a_1^2 a_2^2}{4}r^2\cos(2\theta)}}{r\sin\theta}$ (as defined in Sec. III, $a = a_1^2 a_2$, $\psi_0$ is a constant). A real $v_\varphi$ requires $\psi_0 - \frac{a_1^2 a_2^2}{4}r^2\cos(2\theta) \geq 0$. When $\psi_0 < 0$, this requirement is satisfied outside of the revolution-solid defined by hyperboloid $r^2\cos(2\theta) = \frac{4\psi_0}{a_1^2 a_2^2}$. This is the domain the flow exists (as shown in Fig. 2). Velocity has no singularity in this domain. On the surface of the hyperboloid, $\psi$ is a constant, $v_\varphi$ is zero, and velocity is tangent to the surface.

The requirement of $\psi_0 < 0$ is critical to ensure feasibility of the flow. If $\psi_0 > 0$, the hyperboloid has two sheets. The domain [in which $\psi_0 - \frac{a_1^2 a_2^2}{4}r^2\cos(2\theta) \geq 0$] intersects z-axis. In that z-axis segment, $v_\theta$ and $v_\varphi$ are singular and the flow is not feasible.

Similar to the spherical coordinate case, $H_1C_2$ in cylindrical coordinates

$$\frac{\partial^2\psi}{\partial\rho^2} - \frac{1}{\rho}\frac{\partial\psi}{\partial\rho} + \frac{\partial^2\psi}{\partial z^2} = -\frac{a}{2} \qquad (29)$$

can be solved by adding a particular solution $\psi^* = -\frac{a}{4}z^2$ or $\psi^* = \frac{a}{8}\rho^2(1 - 2\ln\rho)$ or their combination $\psi^* = k_0\frac{a}{8}\rho^2(1 - 2\ln\rho) - (1 - k_0)\frac{a}{4}z^2$ to Eq. (26).

## VII. BELTRAMI SPHERICAL VORTEX ($H_1C_3$)

As discussed in Sec. IV, for $H_1C_3$, as $\boldsymbol{\omega} = a\boldsymbol{v}$, the velocity field is a linear Beltrami field. The velocity in this case is an eigenvector of the curl operator (and $a$ is the eigenvalue).

In Refs. 21 or 22, a solution of velocity field for this case is given by directly solving $\boldsymbol{\omega} = \boldsymbol{v}$ with variable separation method (i.e., without employing stream function). As flows in this case have some significant properties and are important for further discussion, here we will re-solve it in stream function form.

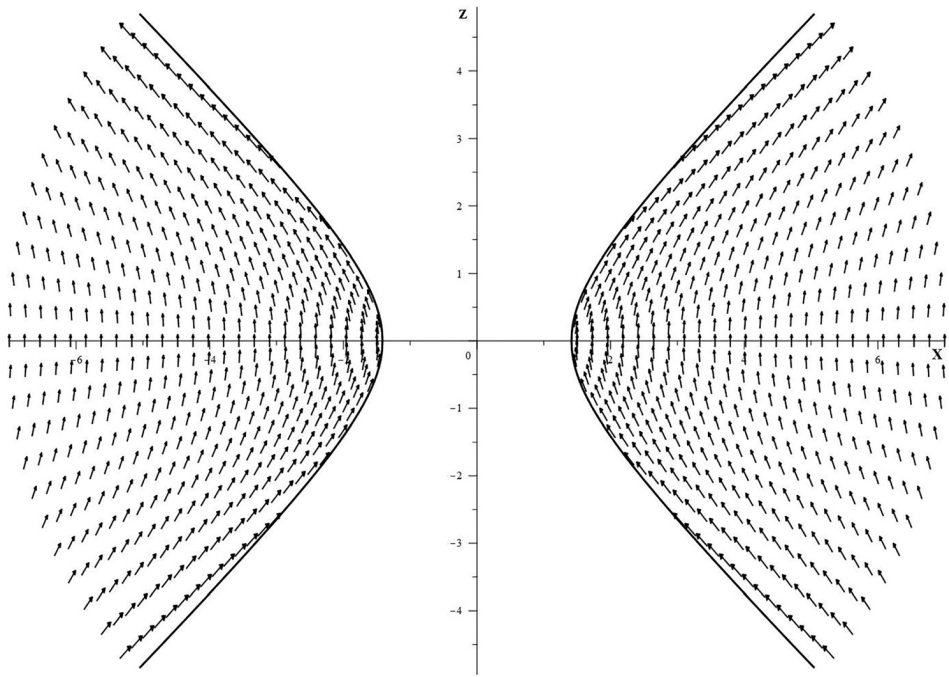

FIG. 2. An $H_1C_2$ flow outside a hyperboloid.

General case of $C_3$ is $C = \alpha\psi + \psi_0$. For convenience, consider $\psi_0 = 0$ first. Equation (14) in this case is

$$\frac{\partial^2 \psi}{\partial r^2} + \frac{\sin^2\theta}{r^2}\frac{\partial^2 \psi}{\partial (\cos\theta)^2} = -a^2\psi. \quad (30)$$

Let $\psi = R(r)\Theta(\cos\theta)$, Eq. (30) separates to the following two equations [with $n(n+1)$ being the separation constant]:

$$r^2\ddot{R} + [a^2 r^2 - n(n+1)]R = 0, \quad (31)$$

$$\sin^2\theta\,\ddot{\Theta} + n(n+1)\Theta = 0. \quad (32)$$

Equation (31) can be converted to a spherical Bessel equation of $F(x)$ if we replace $R(r)$ by $xF(x)$ (with $x = ar$). Thus, it is solved by linear combination of the spherical Bessel function $J_{n+1/2}(ar)$ and $Y_{n+1/2}(ar)$. Equation (32) is the same as Eq. (22) and hence is solved by Eq. (23). Combining these gives the general solution of Eq. (30) as

$$\psi = r^{1/2}\big[k_1 J_{n+1/2}(ar) + k_2 Y_{n+1/2}(ar)\big]$$
$$\times \sin\theta\big[k_3 P_n^1(\cos\theta) + k_4 Q_n^1(\cos\theta)\big]. \quad (33)$$

If we set $a = 1$, $k_2 = 0$, and $k_3 = 1$, $k_4 = 0$, consider the case when $n$ is a positive integer, and re-denote constant $k_1$ as $-K_n$, we have the special case of Eq. (33) as

$$\psi = -K_n r^{1/2} J_{n+1/2}(r)\sin\theta P_n^1(\cos\theta). \quad (34)$$

Submitting this into Eq. (3), velocity can be found the same as in Refs. 21 and 22,

$$\begin{cases} v_r = K_n n(n+1) r^{-3/2} J_{n+1/2}(r) P_n(\cos\theta), & (35a) \\ v_\theta = K_n r^{-1/2}\big[J_{n-1/2}(r) - n J_{n+1/2}(r)/r\big] P_n^1(\cos\theta), & (35b) \\ v_\varphi = -K_n r^{-1/2} J_{n+1/2}(r) P_n^1(\cos\theta). & (35c) \end{cases}$$

This is a family of multi-layer spherical vortices, indexed by $n$ (and will be referred to as Beltrami spherical vortices in this paper). As described in Ref. 22, the field is split by zeros of $J_{n+1/2}(r)$ (i.e., zeros of $v_r$ and $v_\varphi$) into homocentric spherical layers. Inside each layer, there are $n$ count of vortex rings, separated by the surfaces $P_n^1(\cos\theta) = 0$ (i.e., $v_\theta = 0$, $v_\phi = 0$). Figure 3 shows contours of the stream function when $n$ is 1, 2, and 6. Examples of velocity field (when $n = 1, 2, 3$) can also be found in Sec. 15 of Ref. 22.

As shown by the contours, the zero surfaces of $J_{n+1/2}(r)$ and $P_n^1(\cos\theta)$ (i.e., zero surfaces of $\psi$) split the whole space into axisymmetric and coaxial cells that each contains one vortex ring. On these zero surfaces, velocity is only on the tangential direction. Hence, the fluid inside each cell is contained within that cell all the time.

From Eqs. (34) and (35), stream function and velocity have no singularity in the whole space. Thus, such a Beltrami vortex theoretically can steadily exist by itself in the whole space. In other words, it does not have to be moving in a potential flow (like Hill's vortex) or to be surrounded by other flow outside a restricted domain to be physically feasible.

Nevertheless, it is also possible to assemble a Beltrami vortex with potential flow outside a sphere, similar to the case of Hill's vortex.

For the first order vortices in the family (i.e., $n = 1$), at the spherical interfaces between layers, where $J_{n+1/2}(r) = 0$, thus $\psi = 0$, both $v_r$ and $v_\varphi$ vanish, and the polar velocity $v_\theta$ is proportional to $P_1^1(\cos\theta) = -\sin\theta$. This exactly matches the case of irrotational flow past a sphere. Thus, we can have this Beltrami vortex inside the sphere and match it with that irrotational flow outside.

Specifically, in this case, the stream function can be written as

$$\begin{cases} \psi = K_1\left(\dfrac{\sin r}{r} - \cos r\right)\sin^2\theta, \\ \quad (r \leq D,\ D \text{ is one of the solutions of } r = \tan r), & (36a) \\ \psi = \dfrac{1}{3}K_1 \cos D\left(r^2 - \dfrac{D^3}{r}\right)\sin^2\theta\ \ (r \geq D). & (36b) \end{cases}$$

Obviously, in this case, the inside and outside flows both have vanished stream function on the surface of sphere $r = D$, and gradient of stream function is continuous on that surface.

Depending on which solution of $r = \tan r$ parameter $D$ is set to, the vortex inside the sphere can have single or multiple layers. Figure 4 shows two examples of such vortices with 1 layer ($D \approx 4.49$) and 3 layers ($D \approx 10.90$), respectively.

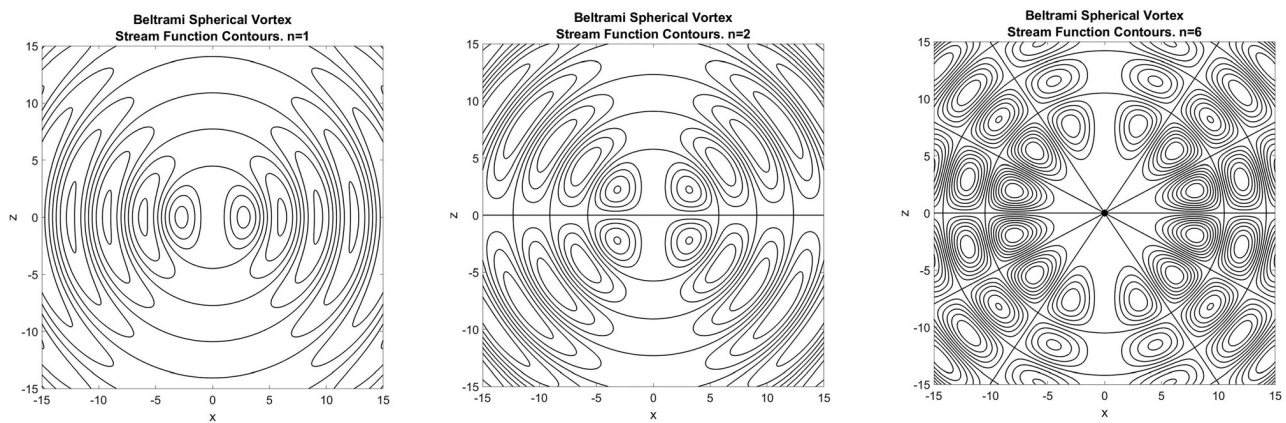

FIG. 3. Stream function contours of some Beltrami spherical vortices (in x–z plane).

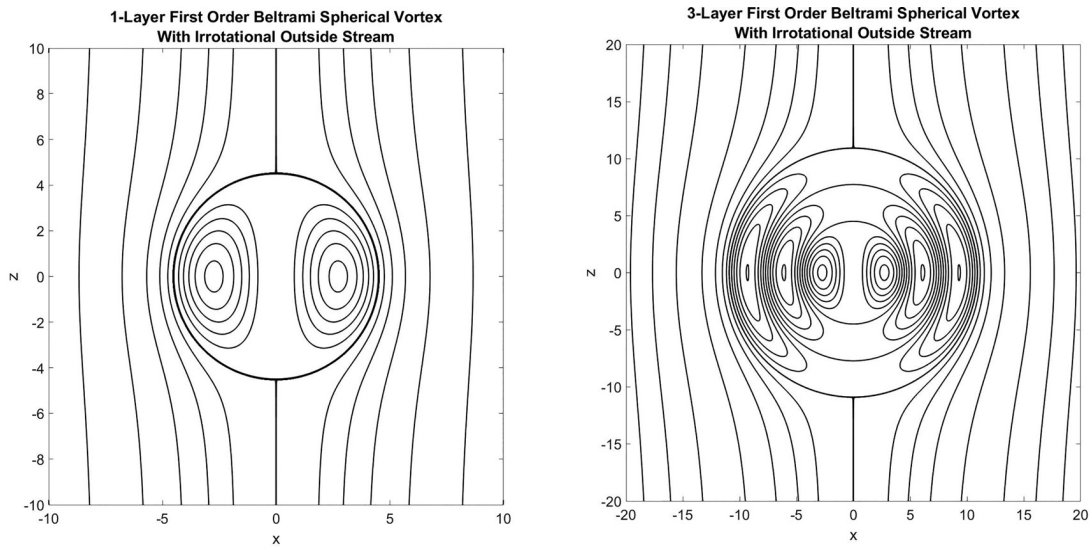

**FIG. 4.** 1- and 3-layer first order Beltrami spherical vortices.

When $n > 1$, $v_r$ and $v_\varphi$ still vanish on the interfaces between spherical layers, but the tangential velocity $v_\theta$ at these interfaces has more complicated dependency to $\theta$ (rather than just being proportional to $\sin\theta$ when $n = 1$). Even in this case, theoretically it is still possible to match the vortex inside a sphere by an (high-order) irrotational flow outside.

Note that for $H_1C_1$, stream function (24) has the same polar angle section as in Eq. (33). With constants properly selected, a high-order (i.e., $n > 1$) axisymmetric potential flow of Eq. (24) outside a sphere can match the same order of Beltrami vortex of Eq. (33) inside that sphere with zero stream function and continuous stream function gradient at the interface. This is an extension of the "assembled flow" of Eq. (36).

In this case, the stream function is

$$\begin{cases} \psi = K_n r^{1/2} J_{n+1/2}(r) \sin\theta P_n^1(\cos\theta), \\ \quad [r \le D, D \text{ is one of the zeros of } J_{n+1/2}(r)], \\ \psi = K_n \dfrac{J_{n-1/2}(D)}{2n+1}(D^{-n+1/2}r^{n+1} - D^{n+3/2}r^{-n}) \\ \quad \times \sin\theta P_n^1(\cos\theta) \ (r \ge D). \end{cases} \quad (37a)$$
$$(37b)$$

When $n = 1$, Eq. (37) simplifies to Eq. (36) (with some constants re-denoted). Two examples of Eq. (37) are shown in Fig. 5. Note that when $n > 1$ the outside flow has unbounded velocity in far field. Such a vortex is more of a theoretical model, as physical feasibility is limited due to the outside flow.

Equation (34) is only a special case of the general solution of Eq. (30). Setting constants in Eq. (33) to other values will give other specific flows.

As a general solution, Eq. (33) [or Eq. (34) as a more specific case] is compatible with many (specific) solutions obtained by other researchers. For example, in Ref. 9, Bogoyavlenskij publishes a solution of z-axisymmetric Beltrami vector field. The $n = 1$ case of Eq. (34) is consistent with that solution [i.e., Eq. (37) in Ref. 9]. In Ref. 4, Bělík et al. present series of axisymmetric Beltrami (Trkalian) flows in different coordination systems. The assumptions for function $H$ and $C$ are equivalent to that of $H_1C_3$ (with $\psi_0 = 0$), and solution (34) is consistent with the spherical coordinate solution obtained in their research (see Sec. 4.2 of Ref. 4).

When $\psi_0 \ne 0$ in $C = \alpha\psi + \psi_0$, the equation for $H_1C_3$ is $\dfrac{\partial^2 \psi}{\partial r^2} + \dfrac{\sin^2\theta}{r^2}\dfrac{\partial^2 \psi}{\partial(\cos\theta)^2} = -a^2\psi + \alpha\psi_0$. This is Eq. (30) plus a constant inhomogeneous term $\alpha\psi_0$ on the RHS. It is solved by adding the particular solution $\psi^* = \dfrac{\psi_0}{\alpha}$ to Eq. (33). (Another way to solve this equation is to replace $\psi$ by $\psi - \dfrac{\psi_0}{\alpha}$ so to convert it to the same form as Eq. (30). This gives the same result.)

In this case, the particular solution $\psi^* = -\dfrac{\psi_0}{\alpha}$ brings to the flow an additional azimuthal velocity, which is singular at z-axis. Such an $H_1C_3$ flow with $\psi_0 \ne 0$ is feasible only when it is inside a domain not overlapping z-axis.

In cylindrical coordinates, when $\psi_0 = 0$, Eq. (15) for $H_1C_3$, is

$$\frac{\partial^2 \psi}{\partial \rho^2} - \frac{1}{\rho}\frac{\partial \psi}{\partial \rho} + \frac{\partial^2 \psi}{\partial z^2} = -a^2\psi. \quad (38)$$

This can be solved the same way as that for Eq. (25). The solution is in the same form as Eq. (26), with constant $k$ in the $P(\rho)$ or in $Z(z)$ section of Eq. (26) biased by $a^2$. That can be written as

$$\psi = \left[k_1 \rho J_1(\sqrt{k}\rho) + k_2 \rho Y_1(\sqrt{k}\rho)\right]\left(k_3 e^{\sqrt{k-a^2}\,z} + k_4 e^{-\sqrt{k-a^2}\,z}\right). \quad (39)$$

Same as in the spherical coordinates case, adding $\psi^* = -\dfrac{\psi_0}{\alpha}$ to Eq. (39) gives solution of the $\psi_0 \ne 0$ case for $H_1C_3$ in cylindrical coordinates.

Mathematically, $H_1C_4$ and $H_1C_3$ have the same form of equation, and thus they share the same solutions. They also have the same $v_r$

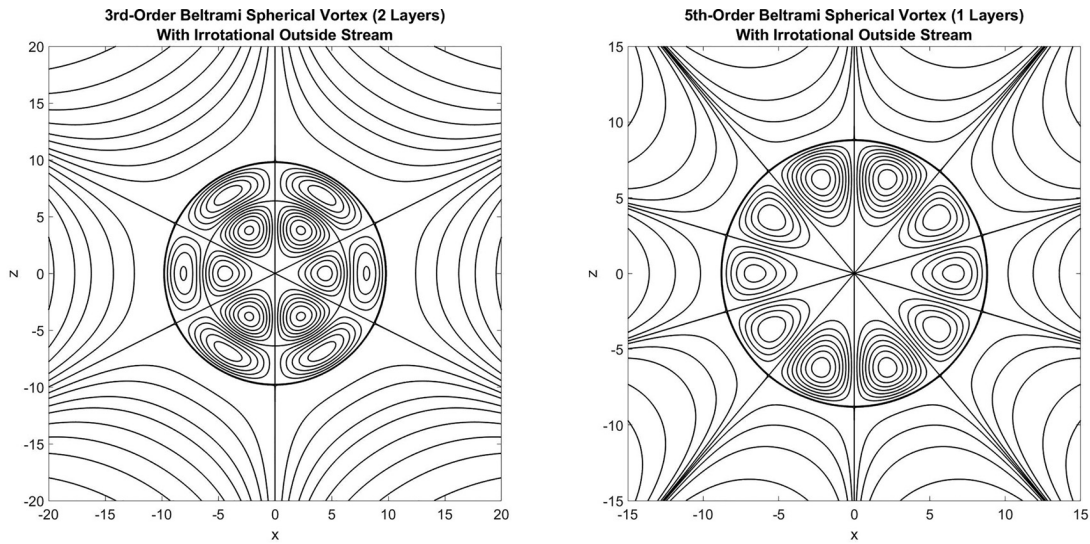

**FIG. 5.** High-order Beltrami spherical vortices surrounded by potential flow.

and $v_\theta$ (or same $v_\rho$ and $v_z$ in cylindrical coordinates), but as these two cases have different function $C$, azimuthal velocity $v_\varphi$ and vorticity are different. As discussed in Sec. IV, the flow in $H_1C_4$ is non-linear Beltrami flow (rather than linear Beltrami flow for $H_1C_3$).

## VIII. HILL'S VORTEX AND EXTENSION ($H_2C_1$)

For Hill's vortex,[7] or a bigger group, the Fraenkel–Norbury family of vortex rings,[10,11,13] the flow in the "core region" is featured by (a) swirl free and (b) vorticity is proportional to distance to the symmetry axis. In terms of B–H equation, feature (a) requires $v_\varphi = 0$, and thus $C = 0$. Considering the vorticity decomposition, vorticity by Eq. (18) becomes $\boldsymbol{\omega} = -r\sin\theta \frac{dH}{d\psi} \boldsymbol{e}_\varphi$ (or $\boldsymbol{\omega} = -\rho \frac{dH}{d\psi} \boldsymbol{e}_\phi$ in cylindrical coordinates). In this case, feature (b) translates to $\frac{dH}{d\psi} = constant$. Flows meeting (a) and (b) hence can be described by $H = H_0 + \lambda\psi$ and $C = 0$. Obviously, this is a subset of $H_2C_1$.

As discussed in Sec. V, for any $H_mC_1$ case, the value of constant $C$ is independent of the equation. Mathematically, a solution in such case is compatible with arbitrary constant $C$. A non-zero $C$ brings non-zero $v_\varphi$ which is an irrotational rotation around the z-axis (and is singular in the z-axis, thus is feasible only in a region not intersecting z-axis). Theoretically speaking (i.e., regardless of the singularity), feature (b) does not have to be paired with feature (a) in a flow.

For any flow of $H_2C_1$, (b) is always valid. In this sense, the inner flow in Fraenkel–Norbury family can be considered as the swirl-free subset of $H_2C_1$ (while Hill's vortex is a special case in the Fraenkel–Norbury family).

Equation (14) for $H_2C_1$ is

$$\frac{\partial^2\psi}{\partial r^2} + \frac{\sin^2\theta}{r^2}\frac{\partial^2\psi}{\partial(\cos\theta)^2} = \lambda r^2 \sin^2\theta. \quad (40)$$

This is Eq. (20) plus an inhomogeneous term on the RHS. A particular solution for Eq. (40) is $\psi^* = \frac{\lambda}{10}r^4\sin^2\theta + \psi_0$. Adding it to the solution of Eq. (20) gives the solution of Eq. (40) as

$$\psi = \left(k_1 r^{n+1} + k_2 r^{-n}\right)\sin\theta \left[k_3 P_n^1(\cos\theta) + k_4 Q_n^1(\cos\theta)\right] + \frac{\lambda}{10}r^4\sin^2\theta + \psi_0. \quad (41)$$

Similar to the case of $H_1C_2$, per Eq. (3), velocity $v_r$ and $v_\theta$ are linear with respect to the stream function $\psi$, thus the particular solution $\psi^*$ in Eq. (41) is linearly adding an extra velocity $\boldsymbol{v}^*$ in meridian plane to the flow of Eq. (24). This extra velocity can be calculated as $v_r^* = \frac{\lambda}{5}r^2\cos\theta$, $v_\theta^* = -\frac{2\lambda}{5}r^2\sin\theta$. Figure 6 shows such a $\boldsymbol{v}^*$ when $\lambda = 1$.

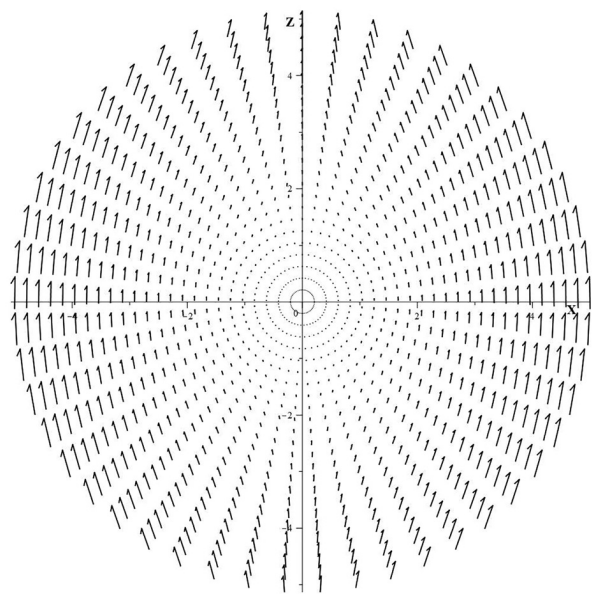

**FIG. 6.** Extra velocity in x–z plane brought by $\psi^*$ for $H_2C_1$ flow.

It is this extra velocity that turns the potential flows of Eq. (24) into $H_2C_1$ flows whose vorticity is proportional to distance to z-axis.

By itself $v^*$ is unbounded in far field. As vorticity is proportional to distance to z-axis, vorticity is also unbounded when $\rho = r\sin\theta$ approaches infinity. To avoid such infinite velocity and vorticity, a physically feasible $H_2C_1$ flow in general should be inside a bounded region, with proper boundary conditions provided by another flow outside (or by a solid boundary). This is the case of the Fraenkel–Norbury family of vortex solutions.

To better discuss, we can consider a more general situation that an $H_2C_1$ flow [i.e., solution (41)] exists inside an axisymmetric domain, $\mathbb{A}$, that (i) on the boundary $\partial\mathbb{A}$ the stream function $\psi$ vanishes, (ii) outside $\mathbb{A}$ (i.e., in domain $R^3 - \mathbb{A}$) exists an axisymmetric flow for which $\psi$ and $\nabla\psi$ are continuous with that of the inside flow at $\partial\mathbb{A}$, and (iii) the inside flow has no singularity in $\mathbb{A}$ and the outside flow has no singularity in $R^3 - \mathbb{A}$ (including $r = \infty$).

Obviously, such an "assembled flow" is a candidate for the Fraenkel–Norbury vortex ring solution. In other words, we can consider a flow meeting (i)/(ii)/(iii) as an extended or generalized Fraenkel–Norbury vortex solution. In this case, the inner flow does not have to be swirl-free, and the outside flow does not have to be vorticity-free. These make it more general than the original Fraenkel–Norbury solutions.

In general, such a flow could be complicated. However, when $n=1$, Eq. (41) can be in a simple form and we have a chance to study the inside flow explicitly.

To avoid singularity, we can further set $k_2 = 0$, $k_3 = 1$, and $k_4 = 0$. With that Eq. (41) becomes

$$\psi = \left(\frac{\lambda}{10}r^4 - k_1 r^2\right)\sin^2\theta + \psi_0. \quad (42)$$

This is a combination of a uniform flow with $U = -2k_1$ in z-direction [i.e., $\psi = -k_1 r^2 \sin^2\theta$ as a special case of Eq. (24)] and the extra velocity $v^*$ as shown in Fig. 6. The polar angle factor, $\sin^2\theta$, is the same as in the axisymmetric potential flow Eq. (24) when also with $k_4 = 0$ and $n = 1$ (i.e., the potential flow past a sphere).

Boundary of the region, $\partial\mathbb{A}$, is defined by $\psi = 0$. This explicitly gives the equation of $\partial\mathbb{A}$ as $r^2\left(k_1 - \frac{\lambda}{10}r^2\right)\sin^2\theta = \psi_0$. This definition is equivalent to the definition by Fraenkel and Norbury in Refs. 10 and 11, which appears as $\psi = k$ ($k$ is a constant). As $\psi_0$ is a free constant in the solution, $\psi_0$ can be set to zero in Eq. (42) and then $\psi$ is a constant on $\partial\mathbb{A}$. In this sense, $\psi_0$ here is equivalent to $-k$ in Refs. 10 or 11.

By this equation, the shape of $\partial\mathbb{A}$ is significantly impacted by $\psi_0$. Without losing generality, assume $\lambda = 1$ and $k_1 = 1$. Figure 7 shows some cross sections of $\partial\mathbb{A}$ with different values of $\psi_0$.

Obviously, these cross sections are consistent in appearance with the numerical results that Norbury presents in Ref. 11.

When $\psi_0 = 0$, $\partial\mathbb{A}$ is the surface of the sphere $r = \sqrt{\frac{10k_1}{\lambda}}$. When $\psi_0 > 0$, $\partial\mathbb{A}$ becomes a "donut shape" inside the sphere. When $\psi_0 = \frac{5k_1^2}{2\lambda}$, $\partial\mathbb{A}$ reaches the extreme case of a "thin core circle" in x–y plane, as defined by $r = \sqrt{\frac{5k_1}{\lambda}}$, $\theta = \frac{\pi}{2}$. The case $\psi_0 < 0$ is not feasible as $\partial\mathbb{A}$ in this case (shown by the dotted line in Fig. 7) encloses the whole z-axis and velocity is unbounded in it.

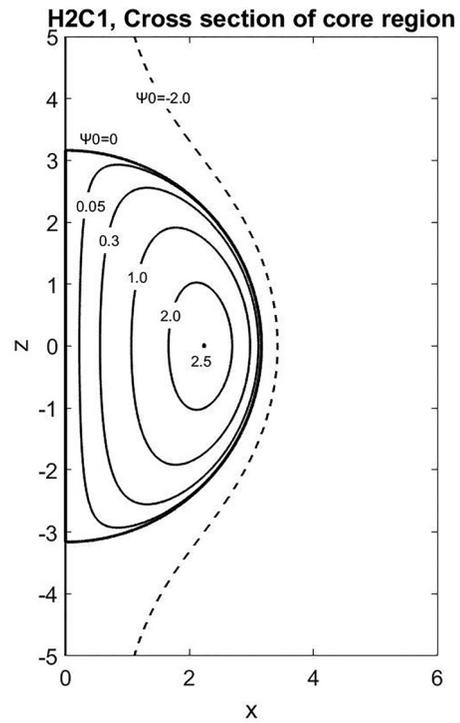

FIG. 7. Cross sections of core region of $H_2C_1$ flow.

When $0 < \psi_0 < \frac{5k_1^2}{2\lambda}$, as the core region is totally avoiding z-axis, the flow inside the core region can have any irrotational rotation (i.e., with any non-zero $C$ and azimuthal velocity $v_\varphi = \frac{C}{r\sin\theta}$). This is a swirled case of the (generalized) Fraenkel–Norbury solutions.

The case of $\psi_0 = 0$ is Hill's vortex. With $\psi_0 = 0$, if we further denote $\lambda = A$ and $k_1 = \frac{A}{10}\alpha^2$, Eq. (42) becomes $\psi = \frac{A}{10}r^2\sin^2\theta(r^2 - \alpha^2)$, which is the well-known stream function of Hill's vortex (in spherical coordinates).

In this case, as a segment of z-axis is involved, there should be no irrotational rotation in the core region (so to avoid singular velocity). In other words, Hill's vortex (as the solution of $H_2C_1$ with $\psi_0 = 0$) is always swirl-free. In comparison, the case of Hill's vortex with swirl, found by Moffat,[8] is actually a solution of $H_2C_3$, which will be discussed in Sec. IX. In that case, total vorticity is no longer proportional to the distance to z-axis, and thus it does not belong to the Fraenkel–Norbury family of vortex rings.

When $\psi_0 > 0$, the outside flow with $\psi$ and $\nabla\psi$ matching the inside flow at the donut-shape boundary could be complicated. However, when $\psi_0 = 0$, as $\partial\mathbb{A}$ is a sphere surface with polar angle factor $\sin^2\theta$ for stream function, a potential flow past that sphere matches well the inside flow. That forms the well-known case of Hill's vortex surrounded by a potential flow.

Recall that the first order Beltrami vortex Eq. (36a) also has polar angle factor $\sin^2\theta$ on the layer interfaces. It is also possible to match Hill's vortex at $\partial\mathbb{A}$ with a "hollow" Beltrami vortex outside. In that case, the stream function can be written as

$$\begin{cases} \psi = K_1\left(\dfrac{\sin r}{r} - \cos r\right)\sin^2\theta, \\ (r \geq D, \; D \text{ is one of the solutions of } r = \tan r), \\ \psi = -K_1\dfrac{\cos D}{2D^2}r^2\sin^2\theta(D^2 - r^2) \; (r \leq D). \end{cases} \quad (43a)(43b)$$

An example of such assembled vortices is shown in Fig. 8 (with $K_1 = 1$ and $D \approx 4.49$). Inside the interface (i.e., sphere $r = D$, shown by the dark line in Fig. 8) is Hill's vortex; outside the interface is the "hollow" Beltrami spherical vortex.

The stream function contours appear to be similar to that of the first order Beltrami spherical vortex in Fig. 3. However, certain properties are quite different inside and outside the interface. Vorticity inside is only in azimuthal direction, with magnitude proportional to distance to the z-axis; vorticity outside is parallel to velocity at each point, with magnitude proportional to magnitude of velocity. On the interface, outside flow has vorticity (and velocity) tangent to the sphere surface (i.e., in the azimuthal direction only). This matches vorticity of inside flow and provides a continuous vorticity at the interface. Another obvious difference between the inside and the outside flows is that one is swirl-free and the other one is with swirl.

Switching to cylindrical coordinates, equation for $H_2C_1$,

$$\frac{\partial^2\psi}{\partial\rho^2} - \frac{1}{\rho}\frac{\partial\psi}{\partial\rho} + \frac{\partial^2\psi}{\partial z^2} = \lambda\rho^2, \quad (44)$$

is the inhomogeneous case of Eq. (25). Adding a particular solution, $\psi^* = \frac{\lambda}{2(4k_5+k_6)}\rho^2(k_5\rho^2 + k_6 z^2) + \psi_0$ to Eq. (26) gives the solution of Eq. (44) as

$$\psi = [k_1\rho J_1(\sqrt{k}\rho) + k_2\rho Y_1(\sqrt{k}\rho)]\left(k_3 e^{\sqrt{k}z} + k_4 e^{-\sqrt{k}z}\right) + \frac{\lambda}{2(4k_5+k_6)}\rho^2(k_5\rho^2 + k_6 z^2) + \psi_0. \quad (45)$$

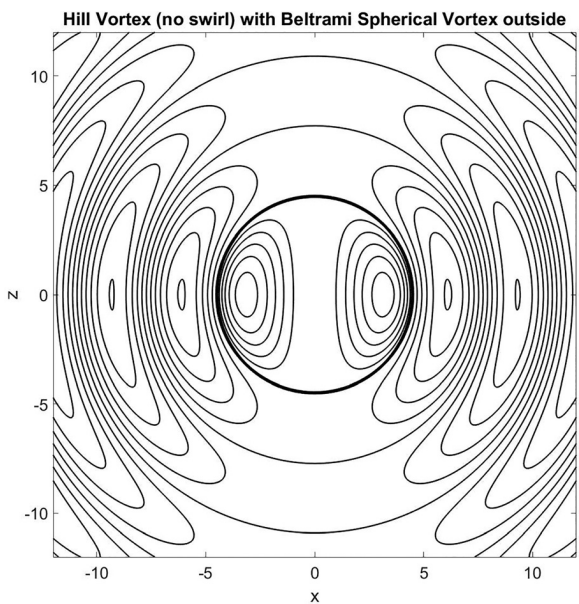

FIG. 8. Hill's vortex (without swirl) with hollow Beltrami spherical vortex.

When $\psi_0 = 0$ and $k_5 = 1$, $k_6 = 1$, the particular solution becomes $\psi^* = \frac{\lambda}{10}\rho^2(\rho^2 + z^2)$. If we consider the special solution of Eq. (25), $\psi = k_1\rho^2$ [see the note in Sec. V following Eq. (26)] and add to it to $\psi^*$, we can also get the Hill's vortex solution $\psi = \frac{A}{10}\rho^2(a^2 - \rho^2 - z^2)$ by re-denoting $\lambda = -A$ and $k_1 = \frac{A}{10}a^2$.

In the case that $k_5$ and $k_6$ are both positive and $k_5 \neq k_6$, $\psi^*$ combined with $\psi = k_1\rho^2$ gives a specific solution $\psi = \frac{A}{10}\rho^2(a^2 - k_5\rho^2 - k_6 z^2)$ (re-denoting $\frac{A}{10} = -\frac{\lambda}{2(4k_5+k_6)}$, $a^2 = \frac{2(4k_5+k_6)k_1}{\lambda}$). This can be considered as an "ellipsoidal shape version" of Hill's vortex.

For $H_2C_2$, another inhomogeneous term, $-\frac{a}{2}$, is added to RHS of the $H_2C_1$ equation, i.e., Eq. (40) or Eq. (44). In spherical coordinates that leads to

$$\frac{\partial^2\psi}{\partial r^2} + \frac{\sin^2\theta}{r^2}\frac{\partial^2\psi}{\partial(\cos\theta)^2} = \lambda r^2\sin^2\theta - \frac{a}{2}, \quad (46)$$

and in cylindrical coordinates, that is,

$$\frac{\partial^2\psi}{\partial\rho^2} - \frac{1}{\rho}\frac{\partial\psi}{\partial\rho} + \frac{\partial^2\psi}{\partial z^2} = \lambda\rho^2 - \frac{a}{2}. \quad (47)$$

These are still inhomogeneous equation of Eqs. (20) and (25), respectively. The particular solutions for Eq. (46) can be found by combining the particular solution of Eq. (27) and the particular solution of Eq. (40). That gives

$$\psi^* = \frac{\lambda}{10}r^4\sin^2\theta - \frac{a}{4}r^2 + \psi_0. \quad (48)$$

Similarly, particular solution for Eq. (47) can be found by combining the particular solution of Eq. (29) and the particular solution of Eq. (44). That yields

$$\psi^* = \frac{\lambda}{2(4k_5+k_6)}\rho^2(k_5\rho^2 + k_6 z^2) + k_0\frac{a}{8}\rho^2(1 - 2\ln\rho) - (1-k_0)\frac{a}{4}z^2 + \psi_0. \quad (49)$$

Adding Eqs. (48) and (49), respectively, to Eqs. (24) and (26) gives the general solutions of Eqs. (46) and (47).

For $H_2C_2$, vorticity is neither proportional to distance to z-axis, nor in parallel and proportional to velocity. According to decomposition equations (18) and (19), vorticity in this case has a non-zero Beltrami vorticity part and a non-zero azimuthal vorticity part. Thus, it is in a combined direction of azimuthal direction and direction of the velocity.

Compared to the $H_1C_2$ or $H_2C_1$ case, the restriction in $H_2C_2$ for velocity and vorticity to be bounded and for $C = a_1\sqrt{\psi_0 + a_2\psi}$ to be real is tighter. A feasible flow in this case would be in a more restricted region compared to that of the $H_1C_2$ or $H_2C_1$ flow.

## IX. HILL'S VORTEX WITH SWIRL ($H_2C_3$)

The Hill's vortex with swirl, discovered by Moffatt[8] and also by Hicks,[15] is a solution of the B–H equation with $C = a\psi$, $H = H_0 + \lambda\psi$. This is the case of $H_2C_3$ (with $\psi_0 = 0$ in $C_3$).

In spherical coordinates, the equation for $H_2C_3$ is

$$\frac{\partial^2 \psi}{\partial r^2} + \frac{\sin^2\theta}{r^2}\frac{\partial^2 \psi}{\partial(\cos\theta)^2} = \lambda r^2 \sin^2\theta - a^2\psi + \alpha\psi_0. \quad (50)$$

Obviously, this is the inhomogeneous case of Eq. (30), i.e., the equation of $H_1C_3$. A particular solution of Eq. (50) is $\psi^* = \frac{\lambda}{a^2}r^2\sin^2\theta + \frac{\psi_0}{\alpha}$. Adding it to Eq. (33) brings the solution of Eq. (50) as

$$\psi = r^{\frac{1}{2}}\left[k_1 J_{n+1/2}(ar) + k_2 Y_{n+1/2}(ar)\right]$$
$$\times \sin\theta\left[k_3 P_n^1(\cos\theta) + k_4 Q_n^1(\cos\theta)\right] + \frac{\lambda}{a^2}r^2\sin^2\theta + \frac{\psi_0}{\alpha}. \quad (51)$$

The same procedure can be employed to solve the $H_2C_3$ equation in cylindrical coordinates,

$$\frac{\partial^2 \psi}{\partial \rho^2} - \frac{1}{\rho}\frac{\partial \psi}{\partial \rho} + \frac{\partial^2 \psi}{\partial z^2} = \lambda\rho^2 - a^2\psi + \alpha\psi_0. \quad (52)$$

With general solution (39) and the particular solution $\psi^* = \frac{\lambda}{a^2}\rho^2 + \frac{\psi_0}{\alpha}$, solution of Eq. (52) is

$$\psi = \left[k_1\rho J_1\left(\sqrt{a^2+k}\rho\right) + k_2\rho Y_1\left(\sqrt{a^2+k}\rho\right)\right]$$
$$\times \left(k_3 e^{\sqrt{k}z} + k_4 e^{-\sqrt{k}z}\right) + \frac{\lambda}{a^2}\rho^2 + \frac{\psi_0}{\alpha}. \quad (53)$$

Alternatively, there is a more straightforward approach to solve Eq. (50). It can be described by the following statement (as a theorem): if $\psi^*$ is solution of Eq. (20), and $\tilde{\psi}$ is solution of Eq. (30), then $\psi = \tilde{\psi} + \psi^*$ is the solution of the following equation:

$$\frac{\partial^2 \psi}{\partial r^2} + \frac{\sin^2\theta}{r^2}\frac{\partial^2 \psi}{\partial(\cos\theta)^2} = a^2\psi^* - a^2\psi. \quad (54)$$

The proof is straightforward. As $\psi^*$ is the solution of Eq. (20), $\psi = \psi^*$ is a particular solution of Eq. (54). While $\tilde{\psi}$ is the solution of Eq. (30), which is the associated homogeneous equation of Eq. (54), $\psi = \tilde{\psi} + \psi^*$ solves Eq. (54).

This theorem is valid for the cylindrical coordinates case as well. In that case $\psi^*$ is the solution of Eq. (25), $\tilde{\psi}$ is the solution of Eq. (38), and $\psi = \tilde{\psi} + \psi^*$ solves the corresponding equation of Eq. (54) in cylindrical coordinates, which is

$$\frac{\partial^2 \psi}{\partial \rho^2} - \frac{1}{\rho}\frac{\partial \psi}{\partial \rho} + \frac{\partial^2 \psi}{\partial z^2} = a^2\psi^* - a^2\psi. \quad (55)$$

Solution of Eqs. (20) and (25) can be in many different forms. Each of them will bring an equation in the form of Eq. (54) or Eq. (55) (not necessarily a B–H equation). Theoretically, this theorem enables an approach to solve these equations.

Equation (50) happens to be a specific case of Eq. (54). When we take $\psi^* = \frac{\lambda}{a^2}r^2\sin^2\theta + \frac{\psi_0}{\alpha}$ as a solution of Eq. (20) and apply the theorem, we obtain Eq. (51). Similarly, Eq. (52) is a specific case of Eq. (55) when we take $\psi^* = \frac{\lambda}{a^2}\rho^2 + \frac{\psi_0}{\alpha}$ as a solution of Eq. (25). Applying the theorem leads to Eq. (53).

As discussed in Sec. VII, solution (33) represents a family of Beltrami spherical vortices. $\psi^*$ in Eq. (51) is adding to them an extra velocity $\mathbf{v}^*$, which by Eq. (3) can be calculated as

$$v_r^* = \frac{2\lambda}{a^2}\cos\theta, \quad v_\theta^* = -\frac{2\lambda}{a^2}\sin\theta, \quad v_\varphi^* = \frac{\lambda}{a^2}r\sin\theta + \frac{\psi_0}{\alpha}\frac{1}{r\sin\theta}. \quad (56)$$

In a meridian plane, $v_r^*$ and $v_\theta^*$ form a constant velocity $U = \frac{2\lambda}{a^2}$ in z direction. On the other hand, azimuthal velocity $v_\varphi^*$ has two terms. The first term $\frac{\lambda}{a^2}r\sin\theta$ is related to a rigid rotation around z-axis; the second term is the irrotational rotation discussed previously, which has singularity on the z-axis. As the flows discussed in this case are mainly vortices centered at $r=0$, to avoid singularity in z-axis, we will only consider the case $\psi_0 = 0$ (i.e., the case without irrotational rotation for $v_\varphi^*$) hereafter in this section.

In this case, the extra velocity $\mathbf{v}^*$ represents a special spiral movement: uniform velocity in z-direction and a rigid rotation around z-axis. Adding this to the linear Beltrami flows of Eq. (33) yields the $H_2C_3$ flows in Eq. (51).

With $\psi_0 = 0$, when $n = 1$, $k_1 = A\alpha^{3/2}$, $k_2 = 0$, $k_3 = -1$, and $k_4 = 0$, Eq. (51) is exactly the inner flow of Hill's vortex with swirl, presented by Moffatt[8] as

$$\psi = r^2\sin^2\theta\left[\frac{\lambda}{a^2} + A\left(\frac{\alpha}{r}\right)^{3/2}J_{3/2}(ar)\right]. \quad (57)$$

Setting constants to other values finds other forms of solution (51).

As nature of rigid rotations, azimuthal velocity $v_\varphi^*$ is unbounded in far field. As a result of that, a feasible $H_2C_3$ flow should be inside a bounded region with proper boundary conditions. When $n=1$ (and $k_4 = 0$), Eq. (51) has a polar angle factor $\sin^2\theta$. This is the same as the irrotational flow past a sphere. Thus, that irrotational flow can provide the boundary conditions.

Taking Eq. (57) as the special case of Eq. (51) and matching it with the irrotational flow outside, with some constants re-denoted, the stream function can be written as

$$\begin{cases} \psi = A\left[\frac{\sin(ar)}{ar} - \cos(ar) + \frac{\lambda}{Aa^2}r^2\right]\sin^2\theta \quad (r \leq D) \\ \left(D \text{ is a positive solution of } \frac{\sin(ar)}{ar} - \cos(ar) \right. \\ \left. + \frac{\lambda}{Aa^2}r^2 = 0 \text{ as an equation of } r\right), \quad (58a) \\ \psi = \left[\frac{\lambda}{a^2} + A\frac{a}{3D}\sin(aD)\right]\left(r^2 - \frac{D^3}{r}\right)\sin^2\theta \quad (r \geq D). \end{cases} \quad (58a)$$

Note that $\frac{\sin(ar)}{ar} - \cos(ar) + \frac{\lambda}{Aa^2}r^2 = 0$ can have multiple positive solutions (depending on value of $\frac{\lambda}{Aa^2}$). Each solution represents a spherical interface of a closed layer. Figure 9(a) shows the flow of Eq. (58a) in the whole space when $A = 1$, $a = 1$, and $\lambda = 0.01$. Two closed layers exist near the center in this case. Outside of the second layer, the flow is no longer contained in close layers, and velocity increases (unboundedly) as $r$ increases.

The flow inside any closed interface can be matched by a potential flow outside. In other words, it does not have to be the first interface from the center. In the case that $D$ in Eq. (58a) is set to a solution other than the first one from the center, the inner vortex has multiple layers. Figure 9(b) shows the central section of Fig. 9(a) with two layers ($D \approx 7.18$) surrounded by potential flow of Eq. (58b).

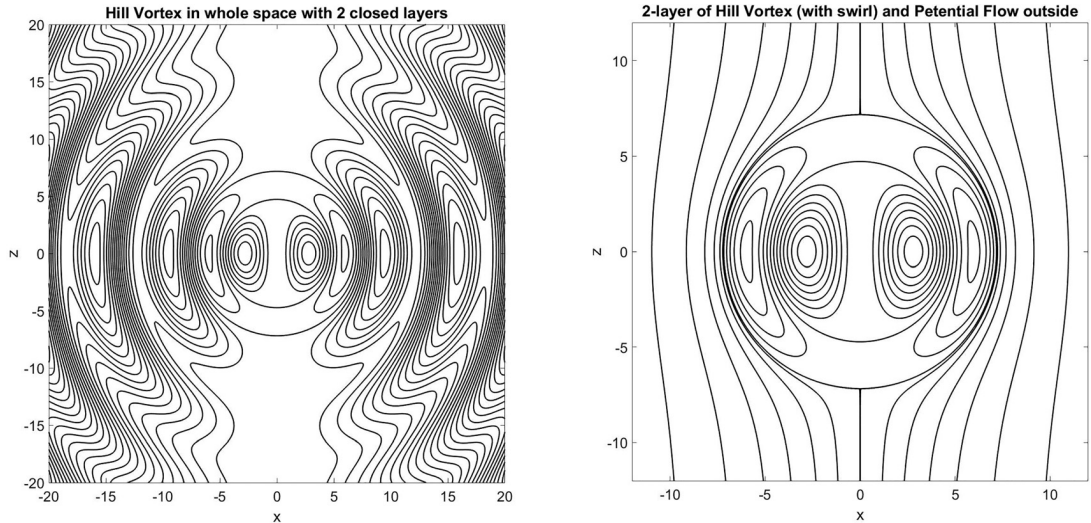

**FIG. 9.** Multi-layer of Hill's vortex with swirl. (a) Hill's vortex with swirl in whole space. (b) 2-layer Hill vortex with swirl in potential flow.

Similar to Hill's vortex without swirl in Eq. (43), the inner vortex with a polar angle factor $\sin^2\theta$ can also be matched by a hollow Beltrami Spherical vertex (similar to that in Fig. 8). In this case, the inner flow is still the same as Eq. (58a), and the outer flow is from Eq. (33) with dedicated constants to have stream function and its gradient matching the inner flow at the interface,

$$\begin{cases} \psi = A\left[\dfrac{\sin(ar)}{ar} - \cos(ar) + \dfrac{\lambda}{Aa^2}r^2\right]\sin^2\theta \ (r \le D) \\ \left(D \text{ is a positive solution of } \dfrac{\sin(ar)}{ar}\right. \\ \left. -\cos(ar) + \dfrac{\lambda}{Aa^2}r^2 = 0 \text{ as an equation of } r\right), \end{cases} \quad (59a)$$

$$\begin{cases} \psi = K\left[\dfrac{\sin(br)}{(br)} - \cos(br)\right]\sin^2\theta \ (r \ge D) \\ \left(K = \dfrac{1}{b\sin(bD)}\left[\dfrac{3\lambda D}{a^2} + aA\sin(aD)\right],\right. \\ \left. b \text{ is a solution of } \dfrac{\sin(bD)}{(bD)} - \cos(bD) = 0\right). \end{cases} \quad (59b)$$

An example of such vortices is shown in Fig. 10 (with $A = 1$, $a = 1$, and $\lambda = 0.01$, $D \approx 7.1798$, $b \approx 1.0760$).

Similar as for $H_1C_4/H_1C_3$, the $H_2C_4$ case has the same equation as $H_2C_3$ and thus shares with $H_2C_3$ the same solution (51) [or solution (53) in cylindrical coordinates]. Feasibility and properties of the $H_2C_4$ flows, however, are quite different from that of $H_2C_3$, even with the same stream function.

## X. OTHER AXISYMMETRIC FLOWS ($H_3C_N$ AND $H_4C_N$)

Some physical background of the $H_3C_n/H_4C_n$ cases can be found in, e.g., Refs. 3, 6, and 23.

For these two families, as $H$ is a quadratic function of $\psi$, the first term on RHS, $r^2\sin^2\theta\dfrac{dH}{d\psi}$ or $\rho^2\dfrac{dH}{d\psi}$, explicitly contains $\psi$. Moreover, in the spherical coordinates case, two coordinate variables, $r$ and $\theta$, are explicitly coupled with $\psi$ in this term. This makes it difficult for variables to be separated. In cylindrical coordinates, however, this term involves only one coordinate variable $\rho$. This gives a chance for variable separation method to apply.

In the rest of this section, we will concentrate on possible solutions for these two families in cylindrical coordinates. The approach to solve them in spherical coordinates is yet to be explored.

For $H_3C_3$ with $\psi_0 = 0$ in $C_3$, Eq. (15) is

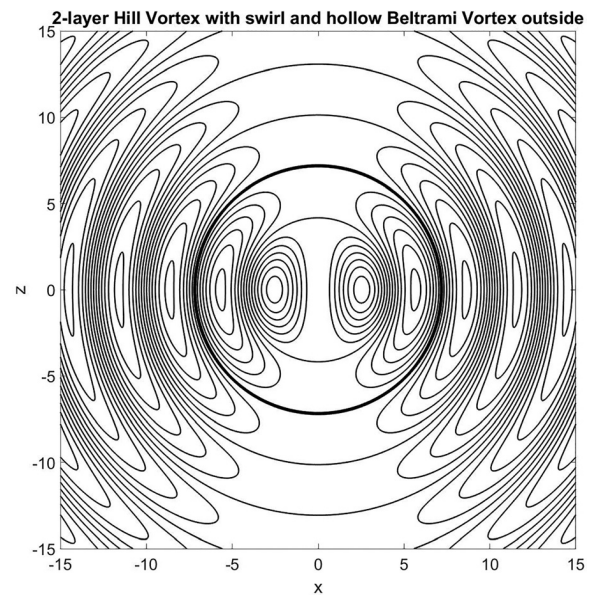

**FIG. 10.** Hill's vortex (with swirl) with hollow Beltrami vortex.

$$\frac{\partial^2 \psi}{\partial \rho^2} - \frac{1}{\rho}\frac{\partial \psi}{\partial \rho} + \frac{\partial^2 \psi}{\partial z^2} = 2\lambda^2\rho^2\psi - a^2\psi. \quad (60)$$

This is the basic homogeneous equation for the two families. We will start the discussion with this equation.

Let $\psi = P(\rho)Z(z)$ and denote the separation constant by $k$, Eq. (60) could be separated to two equations: $\rho\ddot{P} - \dot{P} - (2\lambda^2\rho^2 - k - a^2)\rho P = 0$, $\ddot{Z} - kZ = 0$. The former one can be solved by

$$P(\rho) = \rho^2 e^{-\sqrt{\frac{\lambda}{2}}\rho^2}\left[k_1 M\left(1 - \frac{a^2+k}{4\sqrt{2\lambda}}, 2, \sqrt{2\lambda}\rho^2\right)\right.$$
$$\left. + k_2 U\left(1 - \frac{a^2+k}{4\sqrt{2\lambda}}, 2, \sqrt{2\lambda}\rho^2\right)\right], \quad (61)$$

where $M(a,b,x)$ and $U(a,b,x)$ are Kummer's confluent hypergeometric functions of the first and second kind, respectively.[24]

General solution of Eq. (60) thus is

$$\psi = \rho^2 e^{-\sqrt{\frac{\lambda}{2}}\rho^2}\left[k_1 M\left(1 - \frac{a^2+k}{4\sqrt{2\lambda}}, 2, \sqrt{2\lambda}\rho^2\right)\right.$$
$$\left. + k_2 U\left(1 - \frac{a^2+k}{4\sqrt{2\lambda}}, 2, \sqrt{2\lambda}\rho^2\right)\right]\left(k_3 e^{\sqrt{k}z} + k_4 e^{-\sqrt{k}z}\right). \quad (62)$$

Similar to Eq. (26), in Eq. (62), the section related to z is periodic along z-axis when $k < 0$ (and is unbounded on one side of z-axis when $z \to \pm\infty$ if $k > 0$).

When $a = 0$, Eq. (60) reduces to equation of $H_3C_1$ as

$$\frac{\partial^2 \psi}{\partial \rho^2} - \frac{1}{\rho}\frac{\partial \psi}{\partial \rho} + \frac{\partial^2 \psi}{\partial z^2} = 2\lambda^2\rho^2\psi \quad (63)$$

and Eq. (62) reduces to a solution of Eq. (63) as

$$\psi = \rho^2 e^{-\sqrt{\frac{\lambda}{2}}\rho^2}\left[k_1 M\left(1 - \frac{k}{4\sqrt{2\lambda}}, 2, \sqrt{2\lambda}\rho^2\right)\right.$$
$$\left. + k_2 U\left(1 - \frac{k}{4\sqrt{2\lambda}}, 2, \sqrt{2\lambda}\rho^2\right)\right]\left(k_3 e^{\sqrt{k}z} + k_4 e^{-\sqrt{k}z}\right). \quad (64)$$

For Grad–Shafranov equation in plasma physics, several forms of solutions for equivalent case of Eq. (60) have been published previously. The Herrnegger–Maschke solutions, found by Herrnegger[23] and Maschke,[3] are solving this case by Coulomb wave functions. Cicogna et al. present a solution with confluent hypergeometric function and generalized Laguerre function.[25] Bogoyavlenskij also develops an approach and gains a solution with (integral of) Laguerre polynomials.[6] In Ref. 26, Atanasiu et al. obtain solutions with Coulomb wave functions or in series form with hypergeometric functions. These solutions (including Eq. (62) in this paper) are equivalent but in different formats.

Except for $H_3C_1$, equations for other cases in $H_3C_n/H_4C_n$ families are generally inhomogeneous ones. The corresponding homogeneous equations are in the form of either Eq. (60) or Eq. (63) [with solutions in either Eq. (62) or Eq. (64) forms]. To solve all the cases in these two families, particular solutions for the inhomogeneous equations are needed.

To find the particular solutions, the inhomogeneous cases can be split to two groups. Group A includes $H_3C_2$, $H_4C_1$, and $H_4C_2$, which are homogeneous equation (63) plus one or two inhomogeneous terms. Group B includes the rest cases, $H_3C_3$, $H_3C_4$, $H_4C_3$, and $H_4C_4$, which are homogeneous equation Eq. (60) plus the inhomogeneous term(s).

Recall the model equation (17) in Sec. III, equations for group A can be written in the form of

$$\frac{\partial^2 \psi}{\partial \rho^2} - \frac{1}{\rho}\frac{\partial \psi}{\partial \rho} + \frac{\partial^2 \psi}{\partial z^2} = \Lambda_1\rho^2\psi + \Lambda_0\rho^2 - A_0.$$

If the unknown function $\psi(\rho,z)$ is replaced by $\Phi(\rho,z) - \frac{\Lambda_0}{\Lambda_1}$, the $\Lambda_0\rho^2$ term will be canceled. Given the fact that variable $z$ is not explicitly included on RHS, we can assume that the particular solution is a function of $\rho$ only. In this case, such a particular solution can be found by solving $\psi^*(\rho)$ as a function of $\rho$ in the following ODE (ordinary differential equation):

$$\frac{d^2\psi^*(\rho)}{d\rho^2} - \frac{1}{\rho}\frac{d\psi^*(\rho)}{d\rho} = \Lambda_1\rho^2\psi^*(\rho) - A_0. \quad (65)$$

This gives the particular solution with hyperbolic functions and hyperbolic integrals as

$$\psi^* = [k_1\cosh(\sigma) + k_2\sinh(\sigma)]$$
$$+ \frac{A_0}{2\sqrt{\Lambda_1}}[\text{chi}(\sigma)\sinh(\sigma) - \text{shi}(\sigma)\cosh(\sigma)]$$
$$\left(\sigma = \frac{\sqrt{\Lambda_1}}{2}\rho^2\right). \quad (66)$$

Adding it to Eq. (64) gives the solutions for cases in group A.

For group B, with replacement $\psi(\rho,z) \to \Phi(\rho,z) - \frac{A_0}{A_1}$ or $\psi(\rho,z) \to \Phi(\rho,z) - \frac{\Lambda_0}{\Lambda_1}$, one of the two inhomogeneous terms (but not both) can be eliminated. If either of the following ODE's can be solved, the particular solution is obtained,

$$\frac{d^2\psi^*(\rho)}{d\rho^2} - \frac{1}{\rho}\frac{d\psi^*(\rho)}{d\rho} = \Lambda_1\rho^2\psi^*(\rho) - A_1\psi^*(\rho) - A_0, \quad (67)$$

$$\frac{d^2\psi^*(\rho)}{d\rho^2} - \frac{1}{\rho}\frac{d\psi^*(\rho)}{d\rho} = \Lambda_1\rho^2\psi^*(\rho) - A_1\psi^*(\rho) + \Lambda_0\rho^2. \quad (68)$$

Atanasiu et al. develop an approach to solve equations in Eq. (68) form and obtain the particular solutions in a series form with hypergeometric functions.[26] This is the only solution known by the author of this paper so far. Theoretically, adding that particular solution to Eq. (62) can solve all cases in group B.

## XI. SUMMARY AND DISCUSSION

As discussed in Sec. III, any equation for an $H_mC_n$ case can be written in Eq. (16) or Eq. (17) form. With that, when $\Lambda_0$ and $A_0$ terms both vanish, the equation is homogeneous (and is solved by variable separation mothed). When at least one of $\Lambda_0$ and $A_0$ terms is nonzero, the equation is inhomogeneous and is to be solved by adding a particular solution to solution of the related homogeneous equation. Basically, this is the approach taken in this paper.

The new general solutions obtained in this paper hopefully can benefit the related research in two ways. One is to provide a "high view" for some known vortex solutions to better understand their properties and relations with other flows and to extend them when possible. A good example is Hill's vortex with swirl, which is revealed to be a combination of Beltrami vortex and a special rotation, and can be extended to multi-layer cases. The other way the new general solutions in this paper can contribute is to provide new flows as explicit analytical solutions of B–H equation. For this, the non-linear Beltrami flow discussed in Sec. VI is an example.

TABLE I. Summary of equation forms and solutions.

| H/C function ($H_mC_n$) | Flow property, typical flow | RHS terms $\Lambda_1$ | $\Lambda_0$ | $A_1$ | $A_0$ | Spherical coordinates Equation | Solution | Cylindrical coordinates Equation | Solution |
|---|---|---|---|---|---|---|---|---|---|
| $H_1C_1$ | Irrotational flow | 0 | 0 | 0 | 0 | (20) | (24) | (25) | (26) |
| $H_1C_2$ | Non-linear Beltrami | 0 | 0 | 0 | 1 | (27) | (24) + $\psi^*$ = (28) | (29) | (26) + $\psi^*$ |
| $H_1C_3, \psi_0 = 0$ | Linear Beltrami | 0 | 0 | 1 | 0 | (30) | (33) | (38) | (39) |
| $H_1C_3, \psi_0 \neq 0$ | Linear Beltrami | 0 | 0 | 1 | 1 | (30) + $\alpha\psi_0$ | (33) + $\psi^*$ | (38) | (39) + $\psi^*$ |
| $H_1C_4, \alpha_0 = 0$ | Non-linear Beltrami | 0 | 0 | 1 | 0 | (30) | (33) | (38) | (39) |
| $H_1C_4, \alpha_0 \neq 0$ | Non-linear Beltrami | 0 | 0 | 1 | 1 | (30) + $\alpha\psi_0$ | (33) + $\psi^*$ | (38) | (39) + $\psi^*$ |
| $H_2C_1$ | Hill vortex w/o swirl | 0 | 1 | 0 | 0 | (40) | (24)+$\psi^*$= (41) | (44) | (26)+$\psi^*$= (45) |
| $H_2C_2$ | | 0 | 1 | 0 | 1 | (46) | (24) + (48) | (47) | (26) + (49) |
| $H_2C_3$ | Hill vortex with swirl | 0 | 1 | 1 | 1 | (50) | (33)+$\psi^*$= (51) | (52) | (39)+$\psi^*$= (53) |
| $H_2C_4$ | | 0 | 1 | 1 | 1 | (50) | (33)+$\psi^*$= (51) | (52) | (39)+$\psi^*$= (53) |
| $H_3C_1$ | Homogeneous case | 1 | 0 | 0 | 0 | | | (63) | (64) |
| $H_3C_2$ | | 1 | 0 | 0 | 1 | | | | |
| $H_4C_1$ | Group A | 1 | 1 | 0 | 0 | (yet to be solved) | | (63) + inhomo. term(s) | (64)+$\psi^*$ [$\psi^*$ is from Eq. (66)] |
| $H_4C_2$ | | 1 | 1 | 0 | 1 | | | | |
| $H_3C_3$ | | 1 | 1 | 1 | 1 | | | | |
| $H_3C_4$ | Group B | 1 | 1 | 1 | 1 | (yet to be solved) | | (60) + inhomo. term(s) | (62)+$\psi^*$ ($\psi^*$ is from Ref. 26) |
| $H_4C_3$ | | 1 | 1 | 1 | 1 | | | | |
| $H_4C_4$ | | 1 | 1 | 1 | 1 | | | | |

For convenience, a summary is given in Table I for different cases and the solutions, as well as typical flow properties, when available.

In the table, presentation and absence of the four terms on RHS of the model equation are indicated by "1" and "0" on the "RHS terms" column. $\psi^*$ is representing the particular solution for each inhomogeneous equation. As $\psi_0$ in $C_3$ has big impacts on the equation and solution for $H_1C_3$, cases of $\psi_0 = 0$ and $\psi_0 \neq 0$ are listed separately. Same are the $H_1C_4$ cases with $\alpha_0 = 0$ and $\alpha_0 \neq 0$.

By nature of the variable separation method, solutions can be different when variables are separated in different ways (e.g., in different coordinate systems). Thus, other ways to separate variables may find other solutions. The solutions obtained so far in this paper are not expected to cover all possible solutions.

As the major attention in this paper is on general mathematical solutions, only a few obvious specific flows were discussed as examples. By considering all possible values of the constants, there could be many more specific flows to explore. Physical feasibility (and stability) of them, however, is to be carefully investigated, especially for those "assembled vortices."

## DATA AVAILABILITY

The data that support the findings of this study are available within the article.